\documentclass[letterpaper, 10pt]{article} 
\usepackage[preprint]{tmlr}

\usepackage{amsmath,amsfonts,bm}

\def\eqref#1{equation~\ref{#1}}

\def\1{\bm{1}}

\DeclareMathAlphabet{\mathsfit}{\encodingdefault}{\sfdefault}{m}{sl}
\SetMathAlphabet{\mathsfit}{bold}{\encodingdefault}{\sfdefault}{bx}{n}

\usepackage[utf8]{inputenc} 
\usepackage[T1]{fontenc}    
\usepackage{hyperref}       
\usepackage{url}            
\usepackage{booktabs}       
\usepackage{amsfonts}       
\usepackage{nicefrac}       
\usepackage{svg}
\usepackage{tabularx}
\usepackage{microtype}      
\usepackage{xcolor}         
\usepackage{booktabs}
\usepackage{longtable}
\usepackage{array}
\usepackage{multirow}
\usepackage{graphicx} 
\usepackage{colortbl}
\usepackage{bm}
\usepackage{amsmath}
\usepackage{pifont}
\usepackage{xcolor}
\usepackage{subfigure}
\usepackage{pifont}
\usepackage{cleveref}
\makeatletter
  \renewcommand\p@subfigure{}%
\makeatother

\title{Benchmark on Drug Target Interaction Modeling \\ from a Drug Structure Perspective}

\author{\name Xinnan Zhang* \email zhan9359@umn.edu \\
      \addr University of Minnesota
      \AND
      \name Jialin Wu* \email jlwu@ucsd.edu \\
      \addr University of California San Diego
      \AND
      \name Junyi Xie \email xie00422@umn.edu \\
      \addr University of Minnesota 
      \AND
      \name Tianlong Chen \email tianlong@cs.unc.edu \\
      \addr UNC Chapel Hill
      \AND
      \name Kaixiong Zhou \email kzhou22@ncsu.edu \\
      \addr North Carolina State University }

\def\month{02}  
\def\year{2025} 
\def\openreview{\url{https://openreview.net/forum?id=XXXX}} 

\begin{document}

\maketitle

\begin{abstract}
The prediction modeling of drug-target interactions is crucial to drug discovery and design, which has seen rapid advancements owing to deep learning technologies. Recently developed methods, such as those based on graph neural networks (GNNs) and Transformers, demonstrate exceptional performance across various datasets by effectively extracting structural information. However, the benchmarking of these novel methods often varies significantly in terms of hyperparameter settings and datasets, which limits algorithmic progress. In view of these, we conducted a comprehensive survey and benchmark for drug-target interaction modeling from a structural perspective via integrating tens of explicit (i.e., GNN-based) and implicit (i.e., Transformer-based) structure learning algorithms. We conducted a macroscopical comparison between these two classes of encoding strategies as well as the different featurization techniques that inform molecules' chemical and physical properties. We then carry out the microscopical comparison between all the integrated models across the six datasets via comprehensively benchmarking their effectiveness and efficiency. 
To ensure fairness, we investigate model performance under individually optimized configuration. Remarkably, the summarized insights from the benchmark studies lead to the design of model combos. We demonstrate that our combos can achieve new state-of-the-art performance on various datasets associated with cost-effective memory and computation. 


\renewcommand{\thefootnote}{\fnsymbol{footnote}}
\footnotetext[1]{These authors contributed equally to this work.}
\renewcommand{\thefootnote}{\arabic{footnote}}

\end{abstract}

\section{Introduction}
\label{sec:intro}
The prediction modeling of drug-target interactions (DTI) has emerged as an irreplaceable task for efficacious therapeutic interventions. The binding affinity between a drug molecule and its target protein plays a significant role in the design and repurposing of drugs, where a high affinity typically indicates the desired therapeutics, target specificity, long residence, and drug resistance delay~\citep{hughes2011principles, copeland2006drug, swinney2004biochemical}. The precise modeling of DTI can expedite the drug discovery process and circumvent the associated cost~\citep{Ashburn2004,Strittmatter2014}. 
Deep learning-based frameworks have recently revolutionized this field, enabling more accurate predictions and accelerating the discovery of new compounds by guiding laboratory experiments more efficiently \citep{wen2017deep,abbasi2021deep,huang2020deeppurpose}.

Within deep learning frameworks \citep{ozturk2018deepdta,ozturk2019widedta}, drugs are commonly represented using the Simplified Molecular Input Line Entry System (SMILES)\citep{Smiles}, and proteins are represented as sequences of amino acids. These representations are typically processed using various neural network architectures, such as convolutional neural networks (CNNs)~\citep{krizhevsky2017imagenet, he2016deep}, recurrent neural networks (RNNs), Transformers, and so on, before being integrated and processed by a multi-layer perceptron (MLP) for DTI prediction. It is notorious that the reliance on sequence-based representations can result in the loss of structural information, which can potentially compromise the DTI predictive capability. From the drug perspective, molecular structure modeling helps identify the specific binding sites~\citep{ma2011molecular}, contributes to predicting pharmacokinetic properties~\citep{ekins2007silico}, and allows conformational flexibility~\citep{karplus2005molecular}. 

To address this problem, a number of drug algorithms have been proposed to promote DTI prediction, which can be categorized into explicit and implicit structure learning. First, graph neural networks (GNNs) \citep{kipf2016semi, nguyen2021graphdta} have been widely adopted to learn the molecular structures,
owing to their ability to directly operate on graph-based representations of molecules. By explicitly propagating information through the graph, GNNs can learn node and edge features and thereby capture the structural and functional relationships between atoms and bonds. Second, Transformers, originally focused on natural language processing \citep{Transformer}, have also shown promise in biomedical applications \citep{huang2021moltrans,chen2020Transformercpi}. They rely on self-attention mechanisms to implicitly weight the correlations between different parts of the input SMILES, allowing them to capture long-range dependencies and contextual information. 

While these techniques contribute to the learning of drug structures, there is still a key knob under-explored: we lack a systematic study to benchmark their  effectiveness and efficiency.  
Without such a standardized benchmark, it is unachievable to  offer fair comparisons and subsequently summarize the design philosophy necessary to inform DTI. There have been several surveys and benchmarks on computational methods for DTI prediction\citep{ozturk2018deepdta, huang2020deeppurpose,huang2021therapeutics,xu2022peer}, which leave out the recent developments of structure learning algorithms and unavoidably fail to focus on drug structure benchmarking. Moreover, although massive efforts \citep{bal2024pgraphdta,zhu2023tdgraph,nguyen2021graphdta} have been made to explore the effectiveness of modeling structural information, they predominantly use their proprietary training hyperparameters, datasets, and evaluation metrics. Due to the various settings, one cannot reach convincing answers as to whether a configuration of structure encoders and/or featurization methods generally performs well. The complex of DTI classification and regression tasks and datasets complicates the benchmark comparison.

In this study, we introduce GTB-DTI, a comprehensive \textbf{b}enchmark customized for \textbf{G}NN and \textbf{T}ransformer-based methodologies for \textbf{DTI} prediction. \text{I}) We thoroughly examine the implementation details for each category of drug structure learning methods and integrate three widely used datasets for classification and regression tasks, respectively. We use the optimal hyperparameters reported in their papers to lay the foundation for a fair and reproducible benchmark. 
\text{II}) To gain macroscopical insights into the structure encoders and featurization methods,  we fix the drug encoder to be either GNN or Transformer-based approaches and benchmark these two strategies in the various settings. We 
also integrate tens of drug features given their importance to inform molecules' chemistry and physical properties and evaluate them on the representative datasets. \text{III}) To gain macroscopical insights into nuance between 31 concerned models, we conduct the benchmark studies of their effectiveness on the six datasets. 
Moreover, we assess the efficiency of each method by measuring peak GPU memory usage, running time, and convergences. \text{IV}) The comprehensive study finally provides a number of surprising observations: \textit{i}) The CNN encoder accompanied by integer features has close protein embedding performance compared to the Transformer or larger language models, but they are more efficient. \textit{ii}) The explicit and implicit structure encoders for drugs exhibit unequal performances across the different datasets, which suggests their hybrid usage for generalization purposes. \textit{iii}) Inspired by these insights, we conclude with a model combo that leads us to attaining state-of-the-art (SOTA) regression results and performing similarly to SOTA in the DTI classifications. Our combos further deliver cost-effective memory usage and running time as well as faster convergence, which can serve as a new baseline for the following explorations. 


\section{Formulations for Drug-target Interaction Modeling}

In this research, we focus on the formulations of recently emerging structure modeling approaches for drug molecules, which could be categorized into explicit methods based on graph neural networks and implicit methods based on Transformers. The target proteins are learned by the sophisticated tools of convolutional/recurrent neural networks (CNNs/RNNs) or Transformers, after which both the molecules' and proteins' embeddings are integrated to facilitate interaction prediction. We will also summarize and benchmark the various widely adopted molecule features.

\subsection{Graph Neural Networks based Methods} 
A drug molecule is typically represented as a graph $G=(\mathcal{V},\mathcal{E})$, where $\mathcal{V}$ and $\mathcal{E}$ denote the sets of atoms and chemical bonds, respectively. 
The classical GNN frameworks involve key processes of aggregating and updating node features, collectively referred to as message passing, which can be mathematically represented as follows\citep{scarselli2008graph, duan2022comprehensive}:
\small
\begin{align}
    &\mathbf{h}^{(l+1)}_i = \text{COMBINE}_{\text{node}}^{(l)}\left( \mathbf{h}^{(l)}_i, \text{AGGREGATE}^{(l)}_{\text{node}}\left(f_{\alpha}\left(\left\{ \mathbf{h}^{(l)}_j, \mathbf{e}_{ij}^{(l)}: j \in \mathcal{N}_i \right\} \right) \right) \right), \\
    &\mathbf{e}^{(l+1)}_{ij} = \text{COMBINE}_{\text{edge}}^{(l)}\left( \mathbf{e}^{(l)}_{ij}, \text{AGGREGATE}_{\text{edge}}^{(l)}\left(g_{\beta}\left(\left\{ \mathbf{h}^{(l)}_i,h^{(l)}_j: j \in \mathcal{N}_i \right\} \right) \right) \right),
\end{align}
\normalsize
where $\mathbf{h}_{i}^{(l)}$ is the feature representation of node $v_i$ at layer $l$, $\mathbf{e}_{ij}^{(l)}$ is the feature representation of edge between nodes $v_i$ and $v_j$, $\mathcal{N}_i$ refers to the set of neighboring nodes next to node $v_i$. Functions $\text{AGGREGATE}^{(l)}$, $\text{COMBINE}^{(l)}$ aim to aggregate  the neighborhood representations and integrate them together with the node features, respectively. Additionally, $f_{\alpha}$ and $g_{\beta}$   are feature mapping functions, parameterized by $\alpha$ and $\beta$, respectively. The molecule's representation can be derived using $\text{READOUT}$ a function that processes the set of vertex features $\mathbf{H}^{(L)}$ at the last layer.

\textbf{Graph Convolutional Networks (GCN).}
Given a molecule with $N$ atoms, the adjacency matrix $\mathbf{A} \in R^{N\times N}$ indicates its connectivity, with $A_{ij}=1$ if atom $v_i$ is adjacent to atom $v_j$, and $0$ otherwise. Considering the self-connection of atoms, we have $\tilde{\mathbf{A}} = \mathbf{A} + \mathbf{I}$. Let's $\mathbf{X}\in R^{N\times C}$ denote the initial atom feature matrix. GCN \citep{gcn} models the message passing as follows:
\small
\begin{equation}
\mathbf{H}^{(l+1)}=\sigma(\tilde{\mathbf{D}}^{-\frac{1}{2}}\tilde{\mathbf{A}}\tilde{\mathbf{D}}^{-\frac{1}{2}}\mathbf{H}^{(l)}\mathbf{W}^{(l)}),
\end{equation}
\normalsize
where $\mathbf{H}^{(l)}$ is the node feature matrix at layer $l$, starting with $\mathbf{H}^{(0)} = \mathbf{X}$. Matrix $\mathbf{W}^{(l)}$ represents the learnable weights for layer $l$, $\sigma$ denotes a non-linear activation function, e.g., ReLU, and $\tilde{\mathbf{D}}$ is a diagonal degree matrix of $\tilde{\mathbf{A}}$. 
A couple of pioneering works have leveraged GCN to facilitate drug-protein interaction prediction~\citep{mukherjee2022deepglstm, tran2022deepnc, tsubaki2019compound, pan2023csdti}. For example, DeepGLSTM~\citep{mukherjee2022deepglstm} uses mixture-of-depths GCNs to capture drug representations from different scales. 
CPI~\citep{tsubaki2019compound} considers cross-atom distance and introduces the concept of r-radius subgraphs~\citep{r-radius}, using r-radius vertices and edges to redefine the structure of graphs. 

\textbf{Graph Isomorphism Networks (GIN).}  
GIN excels in learning distinct graph features by approximating the Weisfeiler-Lehman test, enabling it to distinguish a wide range of graph structures~\citep{xu2018powerful}. The message-passing process at the $(l+1)$-th layer is of the following form:
\small
\begin{equation}
\mathbf{h}_{i}^{(l+1)}=\mathrm{MLP}^{(l)}((1+\epsilon^{(l)}) \mathbf{h}_{i}^{(l)}+\sum_{j\in\mathcal{N}_i}\mathbf{h}_{j}^{(l)}),
\end{equation}
\normalsize
where $\mathrm{MLP}^{(l)}$ is a multi-layer perceptron that parameterizes the update function, and $\epsilon^{(l)}$ is a learnable parameter. We benchmark several GIN-based drug-target interaction modeling methods. GraphCPI \citep{quan2019graphcpi} and GraphDTA \citep{nguyen2021graphdta} adopt GIN-based models with batch normalization to obtain the drug representation. SubMDTA \citep{pan2023submdta} uses a subgraph's generation task and contrastive learning to pretrain a molecular graph encoder with multiple GIN layers for further prediction. 

\textbf{Graph Attention Networks (GAT).} Unlike fixed-weight aggregation, GAT~\citep{gat} employs an attention mechanism to determine neighborhood importance and learn the node embeddings as follows:
\small
\begin{equation}
    \mathbf{h}_i^{(l+1)}=\sigma(\sum_{j\in i\cup\mathcal{N}_i}\mathrm{softmax}(\mathrm{LeakyReLU}(\mathbf{W}_a^{\mathrm{T}}[\mathbf{W}^{(l)}\mathbf{h}^{(l)}_i||\mathbf{W}^{(l)}\mathbf{h}^{(l)}_j]))\mathbf{W}^{(l)}\mathbf{h}_j^{(l)}).
\end{equation}
\normalsize
$\mathbf{W}_a^{\mathrm{T}}$ denotes attention weights, and $||$ is a concatenating operation. GraphDTA \citep{nguyen2021graphdta} and AMMVF \citep{wang2023ammvf} leverage the multi-head GAT layers to optimize the atom messaging. They integrate GAT with other architectural modules, such as GCN, facilitating a more comprehensive representation of drugs. 

\textbf{Graph Transformers.} Graph Transformers~\citep{graphTransformer1, maziarka2020molecule} have emerged as powerful alternatives to traditional graph neural networks (GNNs) for molecular representation learning. Unlike conventional GNNs, which rely on message-passing mechanisms to propagate local node information, Graph Transformers leverage self-attention mechanisms to capture both local and global dependencies more effectively. By integrating Message Passing Networks into Transformer-style architectures, these models enhance expressiveness, enabling more comprehensive encoding of molecular structures. This hybrid approach allows Graph Transformers to preserve structural information while benefiting from the flexibility of attention-based learning.

\subsection{Transformer-based Methods} 
Besides the graph representation, drugs could also be decorated as SMILES strings~\citep{weininger1988smiles} and encoded similarly to natural language processing. Specifically, after tokenizing SMILES strings, the Transformer model utilizes multi-head attention to model the interactions between different segments of the input and obtain the molecular representations. Positional encodings are also integrated to preserve the sequence order, enhancing the model's ability to process sequential information effectively. We review and benchmark two typical types of attention mechanisms used for molecular representations.

\textbf{Self-Attention}\citep{huang2021moltrans,qian2023mcl,yin2024fotf}\textbf{.}  Self-attention computes a weighted sum of all input values based on their relevance to each other.  
Considering an embedding of a SMILES sequence $\mathbf{H}^{(l)} \in \mathbb{R}^{d \times N}$ at a specific Transformer layer, where $N$ and $d$ are token length and dimension, respectively, the attention is calculated by $\mathrm{Attention}(\mathbf{Q}, \mathbf{K}, \mathbf{V})=\mathrm{softmax}(\mathbf{Q}\mathbf{K}^{\mathrm{T}} / \sqrt{d_k})\mathbf{V}$. $\mathbf{Q}, \mathbf{K} \in \mathbb{R}^{d_k \times N}$ and $ \mathbf{V} \in \mathbb{R}^{d_v \times N}$ are projections of the input matrix $\mathbf{H}^{(l)}$. 
Multi-head attention combines these projections across different subspaces for a more detailed analysis. Followed by normalization and feed-forward neural networks, the SMILES embedding is updated, $\mathbf{H}^{(l+1)}$ and the output from the last layer is treated as molecular representations. Transformer encoders like MolTrans \citep{huang2021moltrans} and FOTFCPI \citep{yin2024fotf} are adopted to enhance substructure embeddings in proteins and drugs.

\textbf{Cross-Attention}\citep{kurata2022ican, qian2023mcl}\textbf{.} Cross-attention is designed to capture the interaction between the drug and protein sequences, with 
the query matrix $\mathbf{Q}$ derived from one sequence and the key and value matrices $\mathbf{K}, \mathbf{V}$ from another. This mechanism is particularly useful in integrating hybrid representations such as drug graphs and SMILES \citep{wang2023ammvf}, as well as drugs and proteins \citep{pan2023csdti,kurata2022ican}.

\subsection{Feature Processing Methods}

Beyond the drugs' structure or sequence learning with GNNs or Transformers, the extra molecular properties, such as molecular weight, solubility, and lipophilicity, are crucial for building accurate and quantitative drug-target relationship models. We summarize two typical featurization methods.

\textbf{Sequence Processing Methods.} Both drugs and proteins are input as strings of ASCII characters, whose features can be extracted using statistical solutions. Integer encoding \citep{nguyen2021graphdta} simply converts the string to a sequence of integers, which assigns an integer to each character. The N-gram \citep{10.1093/bioinformatics/bti801} captures the statistical dependencies between characters in an input string. Specifically, a 3-gram model breaks down a sequence $S = \{s_1,s_2, ..., s_m\}$ into $\{[s_1,s_2,s_3], [s_2,s_3,s_4],..., [s_{m-2},s_{m-1},s_m]\}$, analyzing the relationship between adjacent characters. 


\textbf{Drug-Unique Featurization Methods.}  
The additional chemical properties and structural details of SMILES strings are often considered to gain a more comprehensive understanding. Extended-Connectivity Fingerprints (ECFP) \citep{doi:10.1021/c160017a018Morgan, doi:10.1021/ci100050tecfp}, involves generating unique identifiers for atoms based on their local chemical environment and iteratively updating these through a hash function to capture a broader molecular context, ultimately producing a set of fingerprints that represent the molecule’s overall structure. Another approach, RDKit, is used to convert SMILES into molecular graphs \citep{landrum2006rdkit, nguyen2021graphdta}, where nodes represent the physical and chemical properties of molecules, and bonds are represented by an adjacency matrix. For example, atomic properties such as atom type, degree, and hydrogen information (like the number of explicit hydrogens) are all crucial for constructing a graph. More detailed properties can be found in Appendix \ref{sec:feat}.

\textbf{Embedding Featurization Methods.} Embedding methods are used to translate these discrete sequences into continuous embedding spaces. Notably, Smi2Vec \citep{8621313Smi2Vec} and Prot2Vec \citep{10.1371/journal.pone.0141287Prot2Vec} convert discrete tokens of drug SMILES and protein sequences into vectors that encapsulate semantic and syntactic similarities, effectively grouping similar tokens together in vector space. 
Additionally, pretrained language models \citep{bal2024pgraphdta,lin2022language} are increasingly utilized to leverage large-scale learned patterns, fine-tuned to analyze complex protein data representations effectively.

\section{A Fair Benchmark Platform Setup}
\label{sec:platform}
\textbf{Benchmark Model and Dataset Selection.} From the perspective of reproducibility, we restrict our analysis to models for which the source code has been publicly released. 
To enhance the comprehensiveness, credibility, and sophistication of our benchmark, we conduct experiments on more than 30 models, including both GNN-based and Transformer-based methods. These models are derived from papers spanning the years 2018 to 2024. 
We run these models on 6 frequently evaluated datasets, including both binary interaction classification and continuous affinity regression. For the classification aspect, we utilize datasets including Human \citep{liu2015improving}, \textit{Caenorhabditis elegans} (\textit{C. elegans}) \citep{tsubaki2019compound}, and DrugBank \citep{wishart2008drugbank}. For regression, we employ the Davis \citep{davis2011comprehensive}, KIBA \citep{tang2014making}, and BindingDB datasets \citep{liu2007bindingdb} with dissociation constant (Kd) measures, as processed in \citet{huang2021therapeutics}. The statistical details of these models and datasets are presented in Appendix \ref{sec:model} and Table \ref{tab:datasets}, respectively. 


\textbf{Hyperparameter Configuration.} 
Given the critical role of hyperparameters in achieving optimal performance, we perform a systematic review of the hyperparameters for all selected models in Appendix \ref{sec:hyper}. To ensure that comparisons across models are equitable, we consider comparing each model using its optimal hyperparameters, as reported in the corresponding perspective papers.

\textbf{Data Split.}
We treat each dataset independently to prevent any information leakage that could arise from training a single model on multiple datasets. For duplicate drug–protein pairs in the regression dataset, only the entry with the maximum affinity score is retained. For duplicates in the classification dataset, all entries are removed if conflicting labels are present; otherwise, a single instance is kept. Following data cleaning, each dataset is split into a training set and a test set. We apply k-fold cross-validation on each training set, where each fold consists of a unique training–validation split, and models trained on different folds are completely independent, thereby eliminating any possibility of cross-fold leakage. The training sets are used to fit the models, the validation sets are used to select the best model during training, and the test sets are for final evaluation.

\textbf{Other Training Details.}
We train on the training set and see its performance on the validation set at the end of every epoch, and the model that achieves the best validation performance will be saved. After training, we evaluate the saved model on the test set and save their results. We average the final performance metrics across all folds as the final results. Considering the original training epochs, we use 1000 as maximum epochs limitation. To avoid overfitting, we consider an early stop mechanism in training. Given the complexity in the dataset, we use 50 patience for all datasets. We use MSE as an early stopping evaluation metric for regression and F1 for classification. The detailed results are provided in Appendix~\ref{sec:expriment}. 

\section{A Macroscopic Benchmark on Encoder and Featurization Strategies}
\label{sec:feat}

\begin{figure}[htbp]
\centering
\includegraphics[scale=0.32]{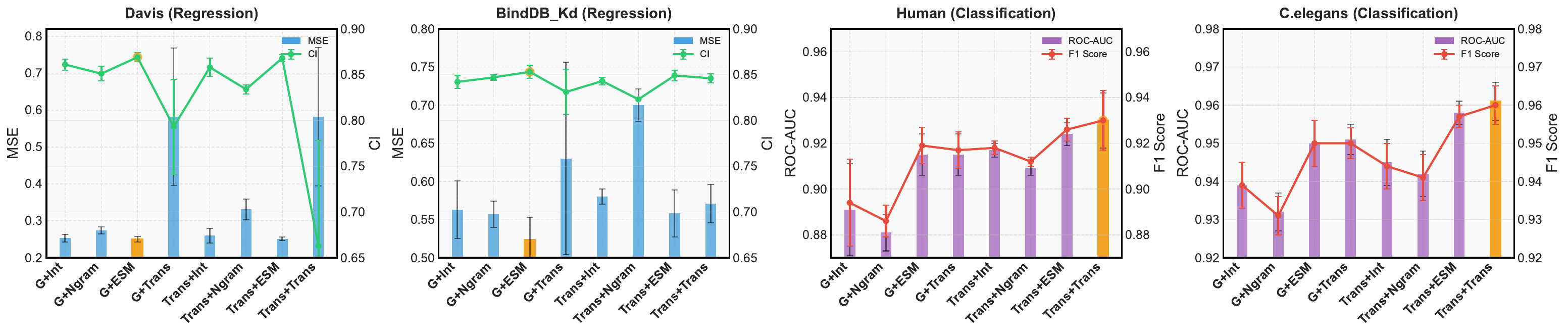}
\caption{Comparison of different encoding strategies with early stop mechanism for drugs and proteins when the total epoch is 1000, LR is 0.0005, BS is 512, and DR is 0.2. Trans is a Transformer-based model, which is composed of two parts: embedding with the position encoding and the encoder in the Transformer. ESM refers to ESM2.}
\label{fig:encoder}
\end{figure}

$\star$\textbf{Encoder Exploration for Drugs and Proteins.}
To investigate the influence of different encoding strategies for extracting the structural information of drugs, we employ GIN~\citep{gin} and vanilla Transformer~\citep{vaswani2017attention} as the encoders for drugs. Meanwhile, integer encoding with CNN, n-gram encoding with CNN, and the vanilla Transformer are considered to capture protein's representations, which are frequently adopted. To leverage the advantages of the pretrained protein information, we include a language model, i.e., Evolutionary Scale Modeling (ESM2) \citep{lin2022language}. The results of various combinations of drug and protein encoders are shown in Fig. \ref{fig:encoder}.All results are averaged by five-fold cross-validation with an early stop mechanism.


\textbf{Obs. 1.} \textbf{GNN and Transformer-based drug encoders exhibit unequal performance depending on DTI tasks.} When the encoder for the protein sequence is fixed, drug features extracted by the GNN structures GIN generally perform better than those by Transformers in regression tasks, but the opposite is true in classification tasks. This disparity may be due to the smaller size of the Human dataset compared to the Davis dataset, which allows for faster convergence in classification tasks than in regression tasks.

\textbf{Obs. 2.}  \textbf{Transformer models are better but sensitive in extracting features from protein.} Although we only consider the simplest pretrained protein language model of ESM2, it still significantly outperforms other encoders in both tasks. This improvement can likely be attributed to the robust and generalizable representations learned from extensive data by the pretrained model. In addition, the Transformer encoder for the protein achieves the best performance on the classification task but shows unstable performance in the regression task. This is likely due to the smaller size and simpler classification dataset compared to the regression dataset, making the training stop for a fixed early stop threshold.

\textbf{Obs 3.} \textbf{Integer encoding appears to be more effective when paired with a CNN as the protein encoder and a fixed drug encoder.} Compared to this specific model configuration, the local context provided by 3-gram encoding does not significantly enhance the model's predictive performance. This implies that the simple relationships in amino acids' immediate neighbors, as modeled by Word2Vec, do not capture much useful information compared with simple integer encoding.

\begin{figure}[htbp]
\centering
\includegraphics[scale=0.3]{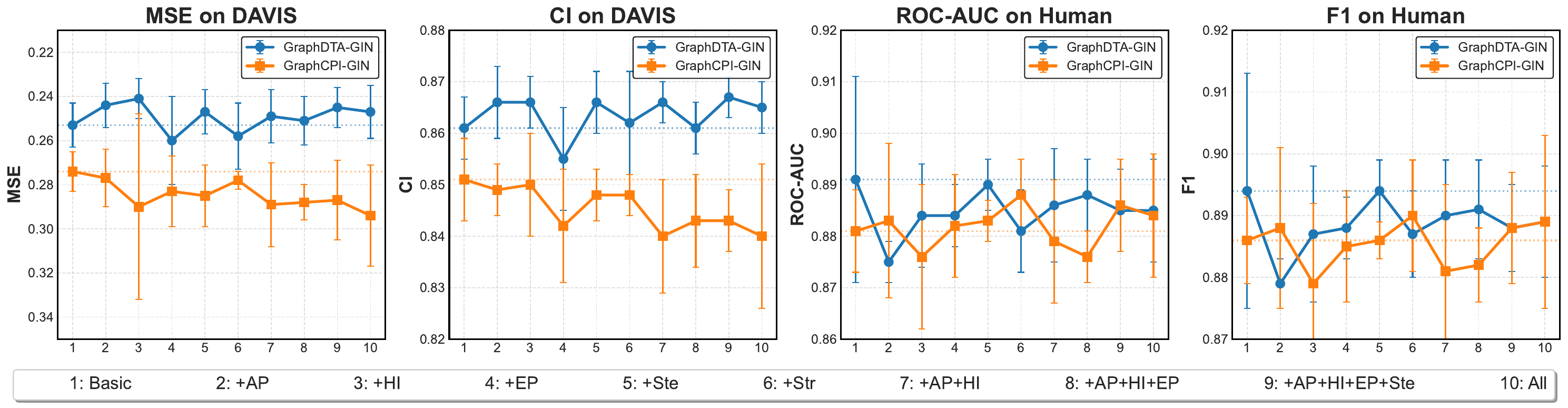}
\caption{Various performances of GraphDTA-GIN and GraphCPI-GIN versus different features on DAVIS and Human datasets. $+ x$ means that $x$ is added to the basic featurization. All means using all features.  }
\label{fig:node_comp}
\end{figure}

$\star$\textbf{Featurization Exploration.}
Despite the efficacy of GNNs in learning drug structures, the featurization of nodes plays a critical role in capturing both the intrinsic properties of atoms and their contextual relevance. We conduct a detailed analysis of various methods (summarized in Section \ref{sec:feat} of the Appendix) for constructing graph features within the DTI context. 
The node feature is constructed via various characteristics, such as chemical and physical properties. We categorize each feature into five main classes, e.g., atomic properties (AP), hydrogen information (HI), electron properties (EP), stereochemistry (Ste), and structural information (Str). To better determine which types of features are more effective in capturing the structural information, we conduct an ablation study on the different featurization strategies. We choose GraphDTA \citep{nguyen2021graphdta} and GraphCPI \citep{quan2019graphcpi} with GIN as our backbone models. The results of feature combinations are reported in Fig.~\ref{fig:node_comp}.


\textbf{Obs. 4.} \textbf{More complex featurization does not necessarily bring a positive effect, and its effectiveness is highly task-dependent.} As shown in  Fig.~\ref{fig:node_comp}, adding features like atomic properties (AP), hydrogen information (HI), and stereochemistry information (Ste) improves performance in the regression task by reducing the MSE loss, suggesting that these features provide valuable information. However, features like electron properties (EP) and structural information (Str) may introduce noise rather than useful information, especially when combined with other features, leading to inconsistent results. Furthermore, in classification tasks, the trend differs, with additional features sometimes negatively impacting performance (blue line) rather than providing benefits. This highlights the importance of careful feature selection, as indiscriminate inclusion of complex features may lead to increased noise, affecting model generalization and robustness.


\textbf{Obs. 5. Protein representation plays a crucial role in the effectiveness of different drug featurizations.} As shown in Fig.~\ref{fig:node_comp}, even when using the same drug featurization strategies, the trends vary depending on the protein representation. This suggests that the way proteins are encoded directly influences how drug features interact with the model. For instance, in GraphDTA, features like atomic properties (AP) and hydrogen information (HI) help improve performance, but in GraphCPI, those same features don’t always provide the same benefits. In some cases, adding more drug features introduces noise rather than useful information. This highlights the fact that drug featurization and protein representation are deeply interconnected, and optimizing one without considering the other may not yield the best results. To build more effective drug-protein interaction models, both components should be considered holistically rather than in isolation. 


\section{A Microscopic Benchmark on DTI Models}\label{section:benchmark}
$\star$\textbf{Benchmark over Effectiveness.}
As shown in Table \ref{tab:regression_bench_s} and Table \ref{tab:classification_bench_s}, we conduct experiments on models with their optimal hyperparameters across two tasks and three datasets, respectively (see more comprehensive model comparisons in Appendix \ref{sec:expriment}). 
All results are averaged by five-fold cross-validation with an early stop mechanism.

\textbf{Obs. 6. Molecular graphs are better than fingerprints for capturing the graph features of a drug.} In reference to Table \ref{tab:featurize_table}, it is evident that GNN-based approaches utilizing the molecular graph generally yield superior performance compared with fingerprints (CPI \citep{tsubaki2019compound}, BACPI \citep{li2022bacpi}, GANDTI \citep{wang2021gandti}). This reinforces the idea that the rich structural and atomic property information inherent to molecular graphs is pivotal for representation extraction, leading to enhanced model performance. 

\textbf{Obs. 7. Graph structure is a crucial part of extracting a drug's features.} Different GNNs have distinct performances in both tasks when the protein representation is fixed. Specifically, GIN, with its unique ability to distinguish non-isomorphic graphs, consistently outperforms other models across different protein encoders in regression tasks. Although Transformer-based methods such as MRBDTA are proficient in handling sequential information from SMILES and proteins, the depth of information they capture appears to be marginally less comprehensive than that provided by molecular graph-based approaches. This is substantiated by the superior performance of GNN-based methods, including MGraphDTA, ColdDTA, and SubMDTA, which suggests that GNN captures intricate structural details more effectively.

\small
\begin{table}[ht]
    \centering
    \caption{Regression task benchmark on DAVIS, KIBA, and BindingDB\_Kd datasets, respectively. For the GraphDTA and GraphCPI, we only show the one with a specific GNN encoder that has the overall best performance. The best result is highlighted in bold, and the runner-up is underlined. \textit{Avg. Reduction of MSE} is computed by the average (across 3 datasets) of the differences between our model's MSE and each model's MSE, divided by the average (across 3 datasets)  of each model's MSE, respectively.}
    \vspace{1em}
    \resizebox{\textwidth}{!}{
    \begin{tabular}{llcccccccccc}
    \toprule
    \multirow{2}{*}{Category} & \multirow{2}{*}{Models} & \multicolumn{3}{c}{\textbf{DAVIS}} & \multicolumn{3}{c}{\textbf{KIBA}} & \multicolumn{3}{c} {\textbf{BindingDB\_Kd}}  & {Avg. Reduction}\\
    \cmidrule(r){3-5}  \cmidrule(lr){6-8} \cmidrule(lr){9-11} & 
    &  MSE & R2 & CI &  MSE  & R2 & CI  & MSE & R2 & CI & of MSE (\%)\\ 
    \midrule
    \multirow{17}{*}{GNN} 
 & GraphDTA-GIN & $0.253 \pm 0.010$ & $0.623 \pm 0.015$ & $0.861 \pm 0.006$ & $0.255 \pm 0.007$ & $-1.840 \pm 0.779$ & $0.553 \pm 0.019$ & $0.563 \pm 0.038$ & $0.693 \pm 0.021$ & $0.842 \pm 0.007$ & $34.827\%$ \\
 & GraphCPI-GIN & $0.274 \pm 0.009$ & $0.593 \pm 0.013$ & $0.851 \pm 0.008$ & $1.681 \pm 0.946$ & $-17.724 \pm 10.533$ & $0.553 \pm 0.094$ & $0.557 \pm 0.017$ & $0.696 \pm 0.009$ & $0.847 \pm 0.003$ & $72.213\%$ \\
 & MGraphDTA & $0.232 \pm 0.012$ & $0.655 \pm 0.018$ & $0.869 \pm 0.007$ & $0.032 \pm 0.012$ & $0.642 \pm 0.133$ & $0.832 \pm 0.040$ & $0.529 \pm 0.011$ & $0.712 \pm 0.006$ & $0.852 \pm 0.005$ & $11.980\%$ \\
 & SAGDTA & $0.324 \pm 0.064$ & $0.518 \pm 0.096$ & $0.833 \pm 0.027$ & $0.065 \pm 0.008$ & $0.279 \pm 0.085$ & $0.713 \pm 0.032$ & $0.529 \pm 0.011$ & $0.712 \pm 0.006$ & $0.852 \pm 0.005$ & $23.965\%$ \\
 & EmbedDTI & $0.280 \pm 0.024$ & $0.583 \pm 0.036$ & $0.851 \pm 0.009$ & $0.289 \pm 0.142$ & $-2.217 \pm 1.579$ & $0.558 \pm 0.038$ & $0.542 \pm 0.019$ & $0.705 \pm 0.010$ & $0.850 \pm 0.004$ & $37.174\%$ \\
 & DeepGLSTM & $0.316 \pm 0.023$ & $0.529 \pm 0.035$ & $0.841 \pm 0.007$ & $8.539 \pm 7.479$ & $-94.109 \pm 83.400$ & $0.514 \pm 0.036$ & $0.594 \pm 0.061$ & $0.677 \pm 0.033$ & $0.840 \pm 0.013$ & $92.613\%$ \\
 & CPI & $0.402 \pm 0.082$ & $0.401 \pm 0.122$ & $0.811 \pm 0.033$ & $0.052 \pm 0.003$ & $0.416 \pm 0.036$ & $0.734 \pm 0.037$ & $0.762 \pm 0.165$ & $0.585 \pm 0.090$ & $0.815 \pm 0.028$ & $42.599\%$ \\
 & BACPI & $0.334 \pm 0.015$ & $0.502 \pm 0.023$ & $0.827 \pm 0.006$ & $0.031 \pm 0.004$ & $0.658 \pm 0.043$ & $0.831 \pm 0.020$ & $0.550 \pm 0.010$ & $0.700 \pm 0.006$ & $0.845 \pm 0.002$ & $23.716\%$ \\
 & DeepNC-HGC & $0.309 \pm 0.025$ & $0.541 \pm 0.037$ & $0.841 \pm 0.005$ & $0.080 \pm 0.003$ & $0.110 \pm 0.036$ & $0.667 \pm 0.022$ & $0.572 \pm 0.011$ & $0.689 \pm 0.006$ & $0.844 \pm 0.003$ & $27.367\%$ \\
 & DeepNC-GEN & $0.270 \pm 0.012$ & $0.597 \pm 0.017$ & $0.852 \pm 0.009$ & $0.135 \pm 0.045$ & $-0.509 \pm 0.505$ & $0.608 \pm 0.037$ & $0.578 \pm 0.020$ & $0.685 \pm 0.011$ & $0.840 \pm 0.003$ & $28.993\%$ \\
 & DrugBAN & $0.242 \pm 0.007$ & $0.640 \pm 0.010$ & $0.869 \pm 0.003$ & $0.029 \pm 0.003$ & $0.676\pm0.032$ & $0.832\pm0.013$ & \underline{$0.465 \pm 0.018$} & \underline{$0.747 \pm 0.010$} & \underline{$0.862 \pm 0.003$} & $5.163\%$ \\
 & GANDTI & $0.318 \pm 0.018$ & $0.527 \pm 0.027$ & $0.844 \pm 0.006$ & $0.030 \pm 0.002$ & $0.662 \pm 0.026$ & $0.831 \pm 0.007$ & $0.621 \pm 0.012$ & $0.662 \pm 0.006$ & $0.836 \pm 0.002$ & $27.967\%$ \\
 & BridgeDPI & $1.241 \pm 1.432$ & $-0.848 \pm 2.133$ & $0.827 \pm 0.078$ & $0.325 \pm 0.109$ & $0.638 \pm 0.121$ & $0.857 \pm 0.001$ & $0.514 \pm 0.011$ & $0.720 \pm 0.006$ & $0.861 \pm 0.002$ & $66.442\%$ \\
 & ColdDTA & \underline{$0.220 \pm 0.009$} & \underline{$0.672 \pm 0.014$} & \underline{$0.880 \pm 0.004$} & $0.110 \pm 0.029$ & $-0.224 \pm 0.329$ & $0.673 \pm 0.079$ & $0.463 \pm 0.008$ & $0.748 \pm 0.004$ & $0.866 \pm 0.001$ & $11.980\%$ \\
 & SubMDTA & $0.289 \pm 0.012$ & $0.570 \pm 0.018$ & $0.841 \pm 0.007$ & \underline{$0.029 \pm 0.002$} & \underline{$0.678 \pm 0.025$} & \underline{$0.836 \pm 0.011$} & $0.532 \pm 0.032$ & $0.710 \pm 0.017$ & $0.852 \pm 0.006$ & $17.882\%$ \\
 & IMAEN & $0.230 \pm 0.009$ & $0.657 \pm 0.014$ & $0.874 \pm 0.004$ & $0.046 \pm 0.018$ & $0.484 \pm 0.196$ & $0.781 \pm 0.056$ & $0.479 \pm 0.012$ & $0.739 \pm 0.006$ & $0.863 \pm 0.002$ & $7.550\%$ \\

\midrule
\multirow{10}{*}{Transformer} 
& CSDTI & $0.331 \pm 0.012$ & $0.508 \pm 0.017$ & $0.832 \pm 0.005$ & $0.088 \pm 0.004$ & $0.014 \pm 0.041$ & $0.628 \pm 0.047$ & $0.768 \pm 0.021$ & $0.582 \pm 0.012$ & $0.805 \pm 0.004$ & $41.196\%$ \\
 & TDGraphDTA & $0.222 \pm 0.005$ & $0.669 \pm 0.008$ & $0.653 \pm 0.011$ & $0.091 \pm 0.019$ & $ -0.009 \pm 0.209 $ & $ 0.327 \pm 0.125 $ & $0.497 \pm 0.016$ & $0.729 \pm 0.009$ & $0.777 \pm 0.005$ & $13.827\%$ \\
 & AMMVF & $0.377 \pm 0.030$ & $0.439 \pm 0.044$ & $0.815 \pm 0.005$ & $0.075 \pm 0.020$ & $0.161\pm0.221$ & $0.603 \pm 0.141$ & $0.682 \pm 0.015$ & $0.628 \pm 0.008$ & $0.825 \pm 0.002$ & $38.448\%$ \\
 & IIFDTI & $0.313 \pm 0.018$ & $0.534 \pm 0.027$ & $0.836 \pm 0.008$ & $0.054 \pm0.013$ & $0.398\pm0.143$ & $0.691 \pm 0.050$ & $0.634 \pm 0.024$ & $0.655 \pm 0.013$ & $0.832 \pm 0.006$ & $30.270\%$ \\
 & ICAN & $0.371 \pm 0.013$ & $0.448 \pm 0.020$ & $0.818 \pm 0.006$ & $0.089 \pm 0.000$ & $-2.052 \pm 0.000$ & $0.500 \pm 0.000$ & $0.747 \pm 0.031$ & $0.593 \pm 0.017$ & $0.813 \pm 0.004$ & $42.171\%$ \\
 & MolTrans & $0.410 \pm 0.136$ & $0.390 \pm 0.202$ & $0.812 \pm 0.039$ & $4.314 \pm 2.290$ & $-47.055 \pm 25.515$ & $0.540 \pm 0.021$ & $0.695 \pm 0.183$ & $0.621 \pm 0.100$ & $0.822 \pm 0.009$ & $87.119\%$ \\
 & TransformerCPI & $0.393 \pm 0.022$ & $0.415 \pm 0.032$ & $0.802 \pm 0.008$ & $0.070 \pm 0.003$ & $0.217 \pm 0.033$ & $0.800 \pm 0.002$ & $0.659 \pm 0.040$ & $0.641 \pm 0.022$ & $0.829 \pm 0.013$ & $37.790\%$ \\
 & MRBDTA & $0.241 \pm 0.005$ & $0.640 \pm 0.008$ & $0.870 \pm 0.007$ & $0.050 \pm 0.005$ & $0.360 \pm 0.058$ & $0.735 \pm 0.015$ & $0.507 \pm 0.006$ & $0.724 \pm 0.003$ & $0.862 \pm 0.002$ & $12.531\%$ \\
  & FOTFCPI & $0.305 \pm 0.012$ & $0.546 \pm 0.018$ & $0.839 \pm 0.009$ & $0.229 \pm 0.180$ & $-1.555 \pm 2.003 $ & $0.587 \pm 0.086$ & $0.567 \pm 0.008$ & $0.695 \pm 0.004$ & $0.848 \pm 0.006$ & $36.603\%$ \\
 & Our combos & \bm{$0.211 \pm 0.007$} & \bm{$0.685 \pm 0.011$} & \bm{$0.886 \pm 0.004$} & \bm{$0.026 \pm 0.004$} & \bm{$0.710 \pm 0.051$} & \bm{$0.849 \pm 0.023$} & \bm{$0.461 \pm 0.006$} & \bm{$0.749 \pm 0.003$} & \bm{$0.869 \pm 0.002$} & $0.000\%$ \\
\bottomrule
\end{tabular}
}
\label{tab:regression_bench_s}
\end{table}
\normalsize

\small
\begin{table}[ht]
\centering
\caption{Classification task benchmark on Human, \textit{C.elegans}, and DrugBank datasets, respectively. For the GraphDTA and GraphCPI, we only show the one that has the overall best performance. The best result is highlighted in bold, and the runner-up is underlined. \textit{Avg. Improvement of Accuracy} is computed by the average (across 3 datasets) of the differences between our model's accuracy and each model's accuracy, divided by the average (across 3 datasets) of each model's accuracy, respectively.}
\vspace{1em}
\resizebox{\textwidth}{!}{
\begin{tabular}{llcccccccccc}
\toprule
\multirow{2}{*}{Category} & \multirow{2}{*}{Models} & \multicolumn{3}{c}{\textbf{Human}} & \multicolumn{3}{c}{\textbf{C.elegans}} & \multicolumn{3}{c}{\textbf{Drugbank}} & Avg. Improvement  \\
    \cmidrule(r){3-5}  \cmidrule(lr){6-8} \cmidrule(lr){9-11} & 
&  ROC-AUC & Accuracy  & F1 &  ROC-AUC & Accuracy & F1 & ROC-AUC & Accuracy & F1 & of Accuracy (\%) \\
\midrule
\multirow{17}{*}{GNN} 
 & GraphDTA-GIN & $0.949\pm0.007$ & $0.885\pm0.011$ & $0.869\pm0.011$ & $0.977\pm0.003$ & $0.929\pm0.005$ & $0.915\pm0.005$ & $0.850\pm0.001$ & $0.783\pm0.006$ & $0.785\pm0.005$ & 3.312\% \\
 & GraphCPI-GIN & $0.941\pm0.005$ & $0.874\pm0.007$ & $0.858\pm0.008$ & $0.971\pm0.003$ & $0.924\pm0.008$ & $0.907\pm0.009$ & $0.838\pm0.012$ & $0.775\pm0.010$ & $0.778\pm0.006$ & 4.275\% \\
 & MGraphDTA & $0.960\pm0.004$ & $0.905\pm0.007$ & $0.893\pm0.007$ & $0.983\pm0.002$ & $0.943\pm0.004$ & $0.931\pm0.004$ & $0.879\pm0.004$ & $0.800\pm0.004$ & $0.806\pm0.003$ & 1.322\% \\
 & SAGDTA & $0.957\pm0.005$ & $0.901\pm0.005$ & $0.887\pm0.006$ & $0.966\pm0.006$ & $0.912\pm0.014$ & $0.894\pm0.017$ & $0.819\pm0.009$ & $0.752\pm0.010$ & $0.756\pm0.010$ & 4.600\% \\
 & EmbedDTI & $0.958\pm0.003$ & $0.901\pm0.005$ & $0.888\pm0.006$ & $0.975\pm0.003$ & $0.924\pm0.002$ & $0.908\pm0.002$ & $0.815\pm0.007$ & $0.758\pm0.005$ & $0.765\pm0.003$ & 3.871\% \\
 & DeepGLSTM & $0.958\pm0.004$ & $0.903\pm0.007$ & $0.890\pm0.008$ & $0.975\pm0.004$ & $0.923\pm0.006$ & $0.906\pm0.007$ & $0.796\pm0.014$ & $0.745\pm0.007$ & $0.752\pm0.006$ & 4.356\% \\
 & CPI & $0.951\pm0.012$ & $0.900\pm0.010$ & $0.887\pm0.012$ & $0.955\pm0.005$ & $0.913\pm0.007$ & $0.893\pm0.010$ & $0.739\pm0.087$ & $0.678\pm0.072$ & $0.687\pm0.074$ & 7.708\% \\
 & BACPI & $0.947\pm0.003$ & $0.905\pm0.007$ & $0.893\pm0.008$ & $0.975\pm0.003$ & $0.936\pm0.005$ & $0.921\pm0.006$ & $0.849\pm0.004$ & $0.776\pm0.009$ & $0.782\pm0.008$ & 2.522\% \\
 & DeepNC-HGC & $0.932\pm0.009$ & $0.861\pm0.015$ & $0.845\pm0.016$ & $0.970\pm0.003$ & $0.918\pm0.004$ & $0.903\pm0.006$ & $0.809\pm0.006$ & $0.752\pm0.006$ & $0.762\pm0.006$ & 6.006\% \\
 & DeepNC-GEN & $0.961\pm0.002$ & $0.907\pm0.006$ & $0.894\pm0.006$ & $0.980\pm0.002$ & $0.932\pm0.005$ & $0.917\pm0.008$ & $0.813\pm0.007$ & $0.736\pm0.015$ & $0.756\pm0.007$ & 4.194\% \\
 & DrugBAN & $0.974\pm0.002$ & $0.920\pm0.005$ & $0.910\pm0.005$ & $0.982\pm0.002$ & $0.946\pm0.004$ & $0.935\pm0.005$ & $0.876\pm0.004$ & $0.799\pm0.008$ & $0.801\pm0.005$ & 0.675\% \\
 & GANDTI & $0.970\pm0.002$ & $0.917\pm0.004$ & $0.906\pm0.004$ & $0.967\pm0.003$ & $0.919\pm0.007$ & $0.901\pm0.007$ & $0.836\pm0.014$ & $0.752\pm0.008$ & $0.763\pm0.004$ & 3.671\% \\
 & BridgeDPI & $0.957\pm0.012$ & $0.887\pm0.021$ & $0.877\pm0.020$ & $0.960\pm0.004$ & $0.882\pm0.034$ & $0.857\pm0.040$ & $0.726\pm0.076$ & $0.644\pm0.087$ & $0.685\pm0.047$ & 11.189\% \\
 & ColdDTA & $0.971\pm0.002$ & $0.922\pm0.009$ & $0.912\pm0.010$ & $0.983\pm0.003$ & $0.947\pm0.002$ & $0.936\pm0.002$ & \bm{$0.885\pm0.004$} & \bm{$0.813\pm0.005$} & \bm{$0.816\pm0.004$} & 0.037\% \\
 & SubMDTA & $0.971\pm0.003$ & $0.919\pm0.006$ & $0.909\pm0.007$ & \underline{$0.985\pm0.001$} & $0.945\pm0.007$ & $0.933\pm0.008$ & $0.861\pm0.005$ & $0.791\pm0.005$ & $0.793\pm0.005$ & 1.055\% \\
 & IMAEN & $0.944\pm0.004$ & $0.878\pm0.005$ & $0.863\pm0.003$ & $0.967\pm0.004$ & $0.911\pm0.007$ & $0.892\pm0.007$ & $0.847\pm0.004$ & $0.777\pm0.005$ & $0.780\pm0.004$ & 4.560\% \\

\midrule
\multirow{10}{*}{Transformer} 
 & CSDTI & $0.905\pm0.007$ & $0.846\pm0.007$ & $0.826\pm0.009$ & $0.910\pm0.006$ & $0.840\pm0.011$ & $0.805\pm0.010$ & $0.774\pm0.011$ & $0.721\pm0.006$ & $0.730\pm0.004$ & 11.467\% \\
 & TDGraphDTA & $0.977\pm0.002$ & $0.927\pm0.005$ & $0.917\pm0.005$ & $0.984\pm0.001$ & $0.943\pm0.007$ & $0.929\pm0.010$ & \underline{$0.880\pm0.006$} & \underline{$0.805\pm0.006$} & \underline{$0.810\pm0.003$} & 0.299\% \\
 & AMMVF & $0.962\pm0.005$ & $0.915\pm0.007$ & $0.905\pm0.009$ & $0.984\pm0.005$ & $0.948\pm0.006$ & $0.937\pm0.007$ & $0.692\pm0.161$ & $0.654\pm0.088$ & $0.696\pm0.020$ & 6.595\% \\
 & IIFDTI & $0.973\pm0.006$ & $0.920\pm0.006$ & $0.909\pm0.008$ & \bm{$0.987\pm0.002$} & $0.948\pm0.005$ & $0.937\pm0.005$ & $0.849\pm0.014$ & $0.777\pm0.010$ & $0.782\pm0.011$ & 1.437\% \\
 & ICAN & $0.971\pm0.002$ & $0.927\pm0.005$ & $0.917\pm0.005$ & $0.977\pm0.004$ & $0.942\pm0.003$ & $0.929\pm0.004$ & $0.839\pm0.005$ & $0.764\pm0.005$ & $0.768\pm0.004$ & 1.899\% \\
 & MolTrans & $0.979\pm0.003$ & $0.931\pm0.002$ & $0.923\pm0.002$ & $0.980\pm0.003$ & $0.943\pm0.004$ & $0.930\pm0.004$ & $0.868\pm0.004$ & $0.795\pm0.005$ & $0.795\pm0.009$ & 0.525\% \\
 & TransformerCPI & $0.968\pm0.003$ & $0.917\pm0.004$ & $0.906\pm0.004$ & $0.984\pm0.001$ & $0.941\pm0.005$ & $0.929\pm0.006$ & $0.874\pm0.007$ & $0.799\pm0.008$ & $0.803\pm0.007$ & 0.979\% \\
 & MRBDTA & $0.971\pm0.004$ & $0.920\pm0.007$ & $0.909\pm0.007$ & \underline{$0.985\pm0.002$} & \underline{$0.953\pm0.002$} & \underline{$0.943\pm0.002$} & $0.866\pm0.005$ & $0.789\pm0.006$ & $0.790\pm0.004$ & 0.789\% \\
 & FOTFCPI & \underline{$0.980\pm0.003$} & \bm{$0.937\pm0.006$} & \bm{$0.929\pm0.006$} & \bm{$0.987\pm0.001$} & \underline{$0.953\pm0.003$} & $0.942\pm0.004$ & $0.866\pm0.002$ & $0.790\pm0.004$ & $0.793\pm0.006$ & 0.112\% \\
 & Our combos & \bm{$0.981\pm0.003$} & \underline{$0.936\pm0.007$} & \underline{$0.928\pm0.008$} & \bm{$0.987\pm0.003$} & \bm{$0.954\pm0.005$} & \bm{$0.944\pm0.006$} & $0.866\pm0.007$ & $0.793\pm0.009$ & $0.798\pm0.005$ & 0.000\% \\

\bottomrule
\end{tabular}
}
\label{tab:classification_bench_s}
\end{table}
\normalsize

\begin{figure}[!htb]
  \centering

  \begin{subfigure}[]{}
    \includegraphics[width=\linewidth]{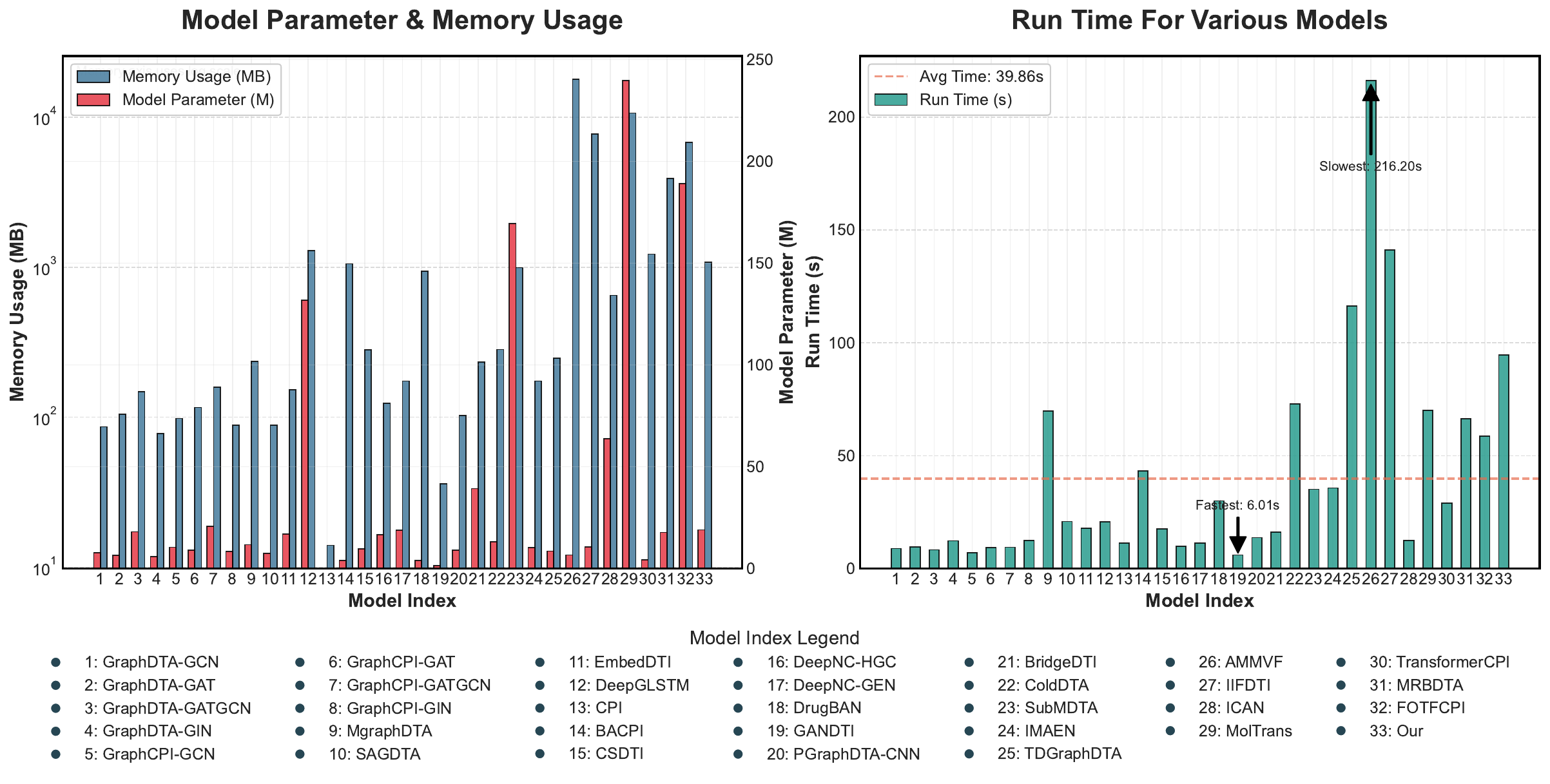}
    \caption{Model size, memory usage and run time on Davis.}
    \label{fig:davis_comp}
  \end{subfigure}

  \vspace{-4pt}

  \begin{subfigure}[]{}
    \includegraphics[width=\linewidth]{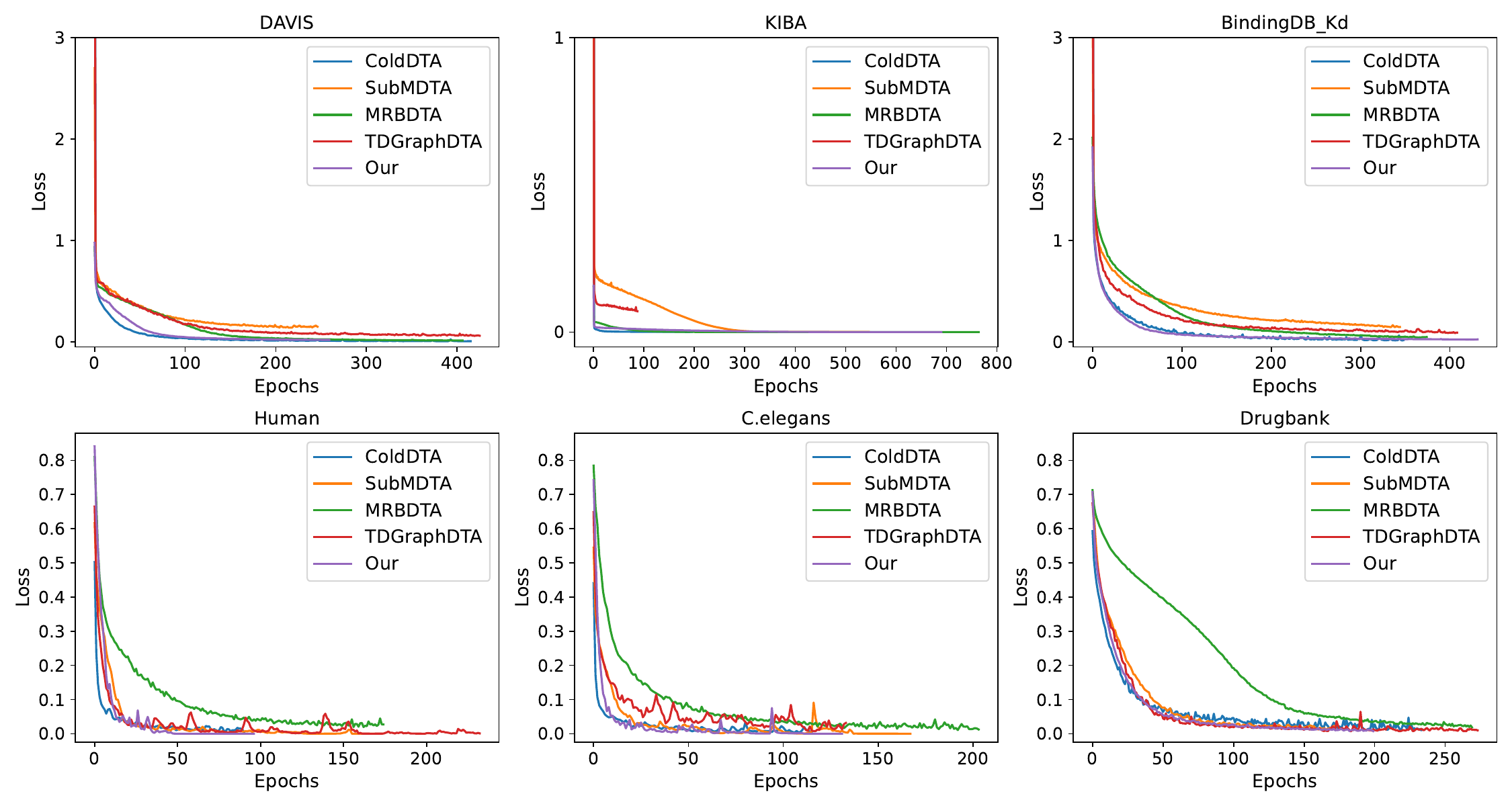}
    \caption{Convergence curves for five selected methods.}
    \label{fig:loss}
  \end{subfigure}

\end{figure}

$\star$\textbf{Benchmark over Efficiency}
To analyze the training speed and memory usage, we empirically evaluate the peak memory and running time for various methods during the training procedure on one regression dataset and one classification task, respectively. To fairly compare various methods, we set the batch size as 32, as such maximum batch size is adopted by some methods. All results are measured on an RTX 3090 GPU. The memory and running time comparisons are illustrated in Fig.~\ref{fig:davis_comp}.

\textbf{Obs. 8. In general, the memory usage of GNN-based methods is smaller than that of Transformer-based methods, which is positively proportional to run time.} This difference is primarily due to the self-attention mechanism employed in Transformers, which requires significant memory resources. In contrast, model parameters, such as those in DeepGLSTM, do not exhibit a direct relationship with either runtime or performance.

$\star$\textbf{Benchmark over Convergence.} We select the two representative methods from the GNN-based and Transformer-based frameworks, respectively, and evaluate them across six datasets on two tasks. The training losses are depicted in Fig. \ref{fig:loss}. To ensure a fair comparison of convergence behavior, we use the previous early stopping setting. Based on the empirical results, we summarize our key observations as follows:

\textbf{Obs. 9. GNN-based methods demonstrate quicker convergence compared to Transformer-based methods.}  This phenomenon arises from the fact that GNN-based methods have less memory usage and fewer model parameters, leading to larger batch size usage or faster convergence compared with Transformer-based methods. 

\subsection{Our Best Combo of Drug and Protein Encoders}
\label{subsection: our}
\textbf{Deriving Combo from Benchmark Insights.} Based on our benchmark results, we summarize the insights of protein and drug encoder usages and propose a light yet effective architecture, which could be treated as a new strong baseline for future explorations. \textbf{\textit{Regarding the proteins}}, 
we observe that 
multi-scale CNNs associated with a mixture of model depths can generally learn the effective protein representations \citep{yang2022MGraphDTA,zhu2023tdgraph,fang2023colddta}, which approximate the language model's accuracy while having lower memory and computation costs. 
\textbf{\textit{Regarding the drug molecules}}, 
both GNN- and Transformer-based methods, such as MRBDTA \citep{zhang2022predicting}, MolTrans~\citep{huang2021moltrans}, and MGraphDTA~\citep{yang2022MGraphDTA} prove promising in DTI tasks. This encourages us to leverage information from hybrid perspectives, i.e., implicit structure (via attention in Transformers) and explicit structure learning (via message passing along edges in GNNs). 

\begin{figure}[htbp]
\centering
\includegraphics[scale=0.35]{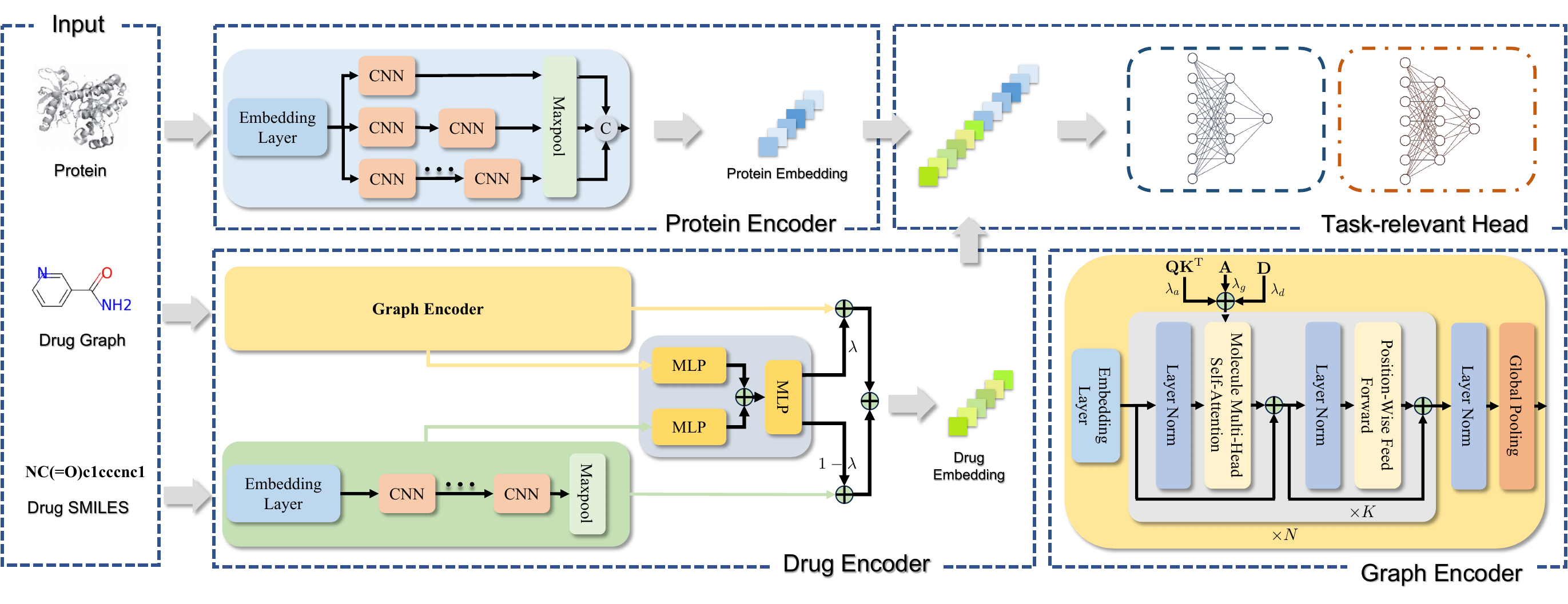}
\caption{Overview of our proposed model combos.}
\label{fig:model}
\end{figure}

Our model design, illustrated in Fig.~\ref{fig:model}, integrates these components. Specifically, for drug graphs, we adopt a hybrid network that augments the self-attention mechanism with inter-atomic distances and graph adjacency matrices \citep{maziarka2020molecule}, incorporating both 2D and 3D molecular structural information.
Given the projections of molecular input at an attention head, i.e.$\mathbf{Q}, \mathbf{K},\mathbf{V}\in \mathbb{R}^{N\times d}$,  the adjacent matrix $\mathbf{A}\in\{0, 1\}^{N\times N}$, and the inter-atomic distances matrix $\mathbf{D} \in \mathbb{R}^{N\times N}$ obtained using RDkit, the augmented attention is calculated as follows:
\begin{align}
    \text{Multi-Attn} = ( \lambda_a \cdot \mathrm{softmax} (\mathbf{Q} \mathbf{K}^{\mathrm{T}} / \sqrt{d})  + \lambda_d g(\mathbf{D}) + \lambda_g \mathbf{A} ) \mathbf{V},
\end{align}
where $g(\cdot)$ is a row-wise softmax function, and $\lambda_a, \lambda_d$ and $\lambda_g$ denote scalars weighting the self-attention, distance, and adjacency matrices, respectively. Besides the implicit and explicit structure learning, we integrate the features from drug SMILES. It is notable that simply utilizing the SMILES representation extracted from a Transformer for downstream tasks does not perform as well as GNN. To align with the protein embedding paradigm, we adopt a simple CNN to unearth potential SMILES information, as suggested in \citet{zhao2022hyperattentiondti}. Subsequently, due to the fact that cross-attention is more complex and hard to optimize, we implement a straightforward attention mechanism to integrate the representations of the drug graph and SMILES, denoted as $\bm{f}_G$ and $\bm{f}_S$, respectively, using a weighting parameter $\lambda$, as follows:
\small
\begin{align}
    \bm{f}_D = \lambda \cdot \bm{f}_G + (1-\lambda) \cdot \bm{f}_S,\ \lambda = \mathrm{MLP}\left(\mathrm{MLP}(\bm{f}_G) + \mathrm{MLP}(\bm{f}_S)\right).
\end{align}
\normalsize

Finally, the prediction is obtained by processing the concatenated protein and drug representations through a task-relevant head, as shown in Fig.~\ref{fig:model}

\textbf{Novelty.} As opposed to the previous strategy, which heuristically stacked a large amount of modules of different types, our model design is driven by systematic benchmarking and empirical insights. Through extensive experiments under fair and controlled conditions, we identify key encoder and featurization strategies that consistently outperform others. Notably, we disentangle and quantify the distinct molecular features. For instance, atomic properties and hydrogen information significantly enhance predictive performance, while adding electron properties may introduce noise.

Thus, our combo is not merely an ad hoc combination but a carefully validated design that effectively obtains a superior balance between accuracy and computational efficiency on both classification and regression tasks. The clear empirical guidance to model design offered through this study helps to establish a more principled framework for future work in drug-target interaction modeling and provides a robust, reproducible new baseline for the community.

\textbf{Benchmark Comparison to State-of-the-Art Frameworks.} We compare the proposed combos with the SOTA frameworks in \Cref{tab:regression_bench_s,tab:classification_bench_s}, and \Cref{fig:davis_comp,fig:loss}. It is observed that our model consistently achieves the best performance in the regression tasks across three datasets and nearly outperforms most methods in classification tasks. By leveraging the physical conformation information from the molecular graph, our combos converge faster than the other two Transformer-based methods, MRBDTA \citep{zhang2022predicting} and TDGraphDTA \citep{zhu2023tdgraph}, particularly on the KIBA dataset. Moreover, our model uses three times less peak memory and fewer parameters than other Transformer-based methods, enabling faster computation and reduced storage requirements.
 
\section{Conclusion}
In this work, we establish a benchmark with fair and consistent experimental configurations, aiming to push DTI research, particularly emphasizing the utilization of structural information. Our meticulous approach has entailed thorough exploration of diverse encoder strategies and featurization techniques for both drug molecules and proteins. Moreover, dozens of existing approaches across six representative datasets for both regression and classification tasks are investigated on various metrics, including DTI classification and regression accuracy, peak memory usage, and model convergence. Provided with the comprehensive benchmark results, we propose a novel approach that integrates the strengths of GNN and Transformer-based methods. Our studies on benchmarking and rethinking help lay a solid, practical, and systematic foundation for the DTI community and provide researchers with broader and deeper insights into the intricate dynamics of drug-target interactions.

\bibliography{main}
\bibliographystyle{tmlr}
\newpage
\appendix

\section{Related Works}

\textbf{GNN-based Methods} GNNs play a crucial role in mining the intricate features of drug molecules for drug-target prediction. Numerous models, including Graph Convolutional Network (GCN), Graph Isomorphism Network (GIN), and Graph Attention Network (GAT), have been utilized \citep{nguyen2021graphdta, quan2019graphcpi,wang2023ammvf, lin2020deepgs, jin2021embeddti} to process and enhance drug features. Additionally, MGraphDTA \citep{yang2022MGraphDTA} employs a multi-scale GNN architecture, while DeepGLSTM \citep{mukherjee2022deepglstm} leverages parallel GNN structures for drug representation. DeepNC integrates advanced techniques from generalized aggregation networks \citep{li2020deepergcn} and hypergraph convolution \citep{BAI2021107637} to improve feature extraction. BACPI \citep{li2022bacpi} develops a bi-directional attention network to integrate the representations of drug molecules and proteins, enhancing their mutual interaction. Besides, BridgeDPI \citep{wu2022bridgedpi} innovates by incorporating bridging nodes between proteins and drugs, utilizing a three-layer GNN for graph embeddings.

\textbf{Transformer-based Methods} Transformers, known for their efficacy in handling sequence data, are extensively applied in drug and protein feature processing. For instance, models like MolTrans \citep{huang2021moltrans} and FOTFCPI \citep{yin2024fotf} employ self-attention mechanisms to refine embeddings by focusing on drug and protein substructures. MRBDTA \citep{zhang2022predicting} uses multi-head attention and skip connection to enhance drug and protein representation. Additionally, a cross-attention mechanism \citep{pan2023csdti, kurata2022ican} is employed to facilitate the integration of drug and protein features, enabling effective mutual querying. TDGraphDTA \citep{zhu2023tdgraph} captures contextual relationships between molecular substructures by using a multi-head cross-attention mechanism and graph optimization. Lastly, DrugormerDTI \citep{hu2023drugormerdti} incorporates degree centrality with positional information to highlight the positional relevance of amino acids in proteins.

\textbf{Input and Featurization} Structural information is crucial at the input stage for models such as BridgeDPI \citep{wu2022bridgedpi}. Various libraries, such as DGLGraph \citep{wang2019DGL}, DGL-lifeSci \citep{li2021dgl-lifesci}, and RDKit \citep{landrum2006rdkit}, are employed to process input SMILES of drugs, with RDKit \citep{landrum2006rdkit} being pivotal for converting SMILE strings into molecular graphs and extracting diverse chemical properties, including chemical bonds, hydrogen presence, electron properties, and so on. Additionally, some approaches \citep{wang2023ammvf,lin2020deepgs,li2022bacpi,wang2021gandti} incorporate molecular fingerprints \citep{doi:10.1021/ci100050tecfp} to capture local chemical information. For protein sequences, typical preprocessing involves converting amino acid sequences into N-grams \citep{pan2023submdta, 10.1093/bioinformatics/bti801} or integer \citep{nguyen2021graphdta} sequences. To enhance the expressiveness of embeddings, some models leverage pre-trained Word2Vec \citep{10.5555/2999792.2999959word2vec, quan2019graphcpi,wang2023ammvf,li2022bacpi,tsubaki2019compound,lin2020deepgs,cheng2022iifdti} or pre-trained protein language models \citep{bal2024pgraphdta}. 

\newpage
\section{Model Descriptions}
\label{sec:model}
This section provides a comprehensive overview of 31 DTI methods, which are classified into GNN-based and Transformer-based approaches. The DTI framework can be simplified as using two encoders to process drugs and proteins separately, followed by an MLP to handle the integrated representations.
\subsection{GNN-based Methods}
\subsubsection{GCN}
$\star$ \textit{GraphDTA-GCN} \citep{nguyen2021graphdta}: GraphDTA-GCN uses GCN to process the molecular graph, which is derived from SMILES using the RDkit tool, and a simple CNN with integer encoding to handle protein sequences.

$\star$ \textit{GraphCPI-GCN} \citep{quan2019graphcpi}: Similar to GraphDTA, GraphCPI-GCN employs 3-gram encoding with pretrained Word2Vec to process protein sequences, followed by a CNN to handle the protein embeddings.

$\star$ \textit{MGraphDTA} \citep{yang2022MGraphDTA}: MGraphDTA utilizes a multiscale GCN, inspired by dense connections, and a multiscale CNN to process drug graphs and protein sequences, respectively.

$\star$ \textit{SAGDTA} \citep{zhang2021sag}: Similar to GraphDTA, SAGDTA introduces global or hierarchical pooling after GCN to aggregate node representations weightedly.

$\star$ \textit{EmbedDTI} \citep{jin2021embeddti}: For protein sequences, EmbedDTI leverages GloVe for pretraining amino acid feature embeddings, which are then fed into a CNN. For drugs, it constructs both an atom graph and a substructure graph to capture structural information at different levels, processed by GCN.

$\star$ \textit{DeepGLSTM} \citep{mukherjee2022deepglstm}: DeepGLSTM processes molecular graphs using a parallel GCN module composed of three GCNs with different layers. For protein sequences, it adopts a bi-LSTM.

$\star$ \textit{CPI} \citep{tsubaki2019compound}: CPI processes drug graphs using GCN. The protein sequence is handled via n-gram with integer encoding, followed by a CNN.

$\star$ \textit{DeepNC} \citep{tran2022deepnc}: DeepNC adopts advanced techniques from generalized aggregation networks and hypergraph convolution, two variants of GCN, to capture the representations of drugs. For protein sequences, it uses a simple CNN.

$\star$ \textit{DrugBAN} \citep{zhang2022predicting}: DrugBAN employs GCN and CNN blocks to encode molecular graphs and proteins, respectively. Then they use a bilinear attention network module to learn local interactions between the representations of drugs and proteins.

$\star$ \textit{BridgeDPI} \citep{wu2022bridgedpi}: BridgeDPI innovates by constructing a learnable drug–protein association network, which is processed using a three-layer GNN for graph embeddings. The learned representations for drug and protein pairs are then concatenated for further processing.

$\star$ \textit{ColdDTA} \citep{fang2023colddta}: ColdDTA removes the subgraphs of drugs. For the model, they adopt the dense GCN and multiscale CNN from MGraphDTA as the encoders for drugs and proteins, respectively. Additionally, an attention-based method is developed to integrate representations for improved prediction.

$\star$ \textit{IMAEN} \citep{zhang2024imaen}: IMAEN employs a molecular augmentation mechanism to enhance molecular structures by fully aggregating molecular node neighborhood information. It then uses multiscale GCN and CNN for drug and protein processing, respectively.

$\star$ \textit{GanDTI} \citep{wang2021gandti}: Inspired by residual networks, GanDTI adds the input drug fingerprints to the output of three GCN layers as graph node features and uses summation to get the final drug representation.

\subsubsection{GAT}

$\star$ \textit{GraphDTA-GAT} \citep{nguyen2021graphdta}: GraphDTA-GAT adopts a GAT as the encoder for drugs, while other components remain the same as in  GraphDTA-GCN.

$\star$ \textit{GraphDTA-GATGCN} \citep{nguyen2021graphdta}: GraphDTA-GATGCN adopts a combination of GAT and GCN as the encoder for drugs, while other components remain the same as in GraphDTA-GCN.

$\star$ \textit{GraphCPI-GAT} \citep{quan2019graphcpi}: GraphDTA-CPI adopts a GAT as the encoder for drugs, while other components remain the same as in  GraphCPI-GCN.

$\star$ \textit{GraphCPI-GATGCN} \citep{quan2019graphcpi}: GraphCPI-GATGCN adopts a combination of GAT and GCN as the encoder for drugs, while other components remain the same as in  GraphCPI-GCN.

$\star$ \textit{BACPI} \citep{li2022bacpi}: BACPI adopts a GAT and a CNN for the features of the fingerprints and protein sequence, respectively. These features are then fed into a bidirectional attention neural network to obtain integrated representations.

$\star$ \textit{PGraphDTA-CNN} \citep{bal2024pgraphdta}: PGraphDTA-CNN is a straightforward method that utilizes GAT for drug feature extraction and CNN for protein sequences.

\subsection{GIN}

$\star$ \textit{GraphDTA-GIN} \citep{nguyen2021graphdta}: GraphDTA-GAT adopts a GAT as the encoder for drugs, while other components remain the same as in GraphDTA-GCN.

$\star$ \textit{GraphCPI-GIN} \citep{quan2019graphcpi}: GraphDTA-GAT adopts a GAT as the encoder for drugs, while other components remain the same as in GraphDTA-GCN.

$\star$ \textit{SubMDTA} \citep{pan2023submdta}: SubMDTA utilizes a pretrained GIN encoder obtained through contrastive learning for the molecular graph. For protein sequences, it employs N-gram embedding with different N to extract features at various scales, which are then processed by a BiLSTM.

\subsection{Transformer-based Methods}

\subsubsection{Self-attention}
$\star$ \textit{AMMVF} \citep{wang2023ammvf}: AWMVF introduces the multi-head mechanism to GAT to learn features in different spaces, and the update function is obtained through the concatenation of different heads' outputs.

$\star$ \textit{IIFDTI} \citep{cheng2022iifdti}: IIFDTI model attains the drug matrix and protein matrix and inputs them to the bi-directional encoder-decoder block, which considers both the drug and target directions. The decoder is mainly composed of multi-head attention. 

$\star$ \textit{MolTrans} \citep{huang2021moltrans}: MolTrans uses Transformer encoder layers to augment the embedding of substructure sequences of proteins and drugs.

$\star$ \textit{FOTFCPI} \citep{yin2024fotf}: Similar to MolTrans, FOTFCPI uses Transformer encoder layers to extract the features of protein and drug fragments after the embedding layers.

$\star$ \textit{TransformerCPI} \citep{chen2020Transformercpi}: TransformerCPI uses the decoder module of Transformer, which takes in the atom sequence embedding processed by GCN and the  protein sequence embedding processed by word2vec and 1D CNN.

$\star$ \textit{MRBDTA} \citep{zhang2022predicting}: In MRBDTA, after the embedding layer, drug sequences are directly fed into a block consisting of three Transformer encoders. The first encoder has a linear layer before it and the following two encoders are parallel. The protein sequence is also processed by a block with a similar structure.

\subsubsection{Cross attention}

$\star$ \textit{CSDTI}\citep{pan2023csdti}: CSDTI uses cross-attention to fuse the deep representations of drugs and proteins. Specifically, the different projections of protein features are used as keys and values, respectively, while the projection of drug features is used as a query. 

$\star$ \textit{TDGraphDTA}\citep{zhu2023tdgraph}: TDGraphDTA uses a multi-head cross-attention mechanism with two attention heads. Both drug and protein features are linearly transformed into query, key, and value matrices. One cross-attention layer uses a drug query matrix, a protein key matrix, and a protein value matrix, while its parallel counterparts use the rest of the matrices. The outputs of these two layers are concatenated and fed into MLP to get the final output.

\newpage
\section{Datasets Descriptions}
In this subsection, we provide a detailed description of the datasets for both the regression task and classification task. The statistical characteristics of the datasets are summarized in Table \ref{tab:datasets}. Here we present the statistics after we cleaned the data as described in Section \ref{sec:platform}.

\begin{table}[htbp]
\centering
\caption{Statistics of the benchmark dataset for two tasks.}
\vspace{1em}
\resizebox{\textwidth}{!}{
\begin{tabular}{lccccccccc}
\hline
  & \multicolumn{3}{c}{\textbf{Regression}} & \multicolumn{3}{c}{\textbf{Classification}} \\ \cmidrule(lr){2-4} \cmidrule(lr){5-7}                  & Davis & KIBA & BindingDB\_Kd    & Human & \textit{C. elegans} & DrugBank \\ \hline
Number of drugs    &     68 & 2068 & 10661      & 2726  & 1767                & 6645     \\
Number of target proteins &  379 & 229 & 1413 &  2001   & 1876                & 4256     \\
Number of total samples   &  25772 & 117657 & 52284        & 5997  & 6552                & 34740    \\ \hline
\end{tabular}
}
\label{tab:datasets}
\end{table}

\begin{figure}[htbp]
    \centering  
    \subfigure[Label distribution of DVAIS, KIBA and Binding\_Kd for regression tasks.]{
    \includegraphics[width=0.4\textwidth]{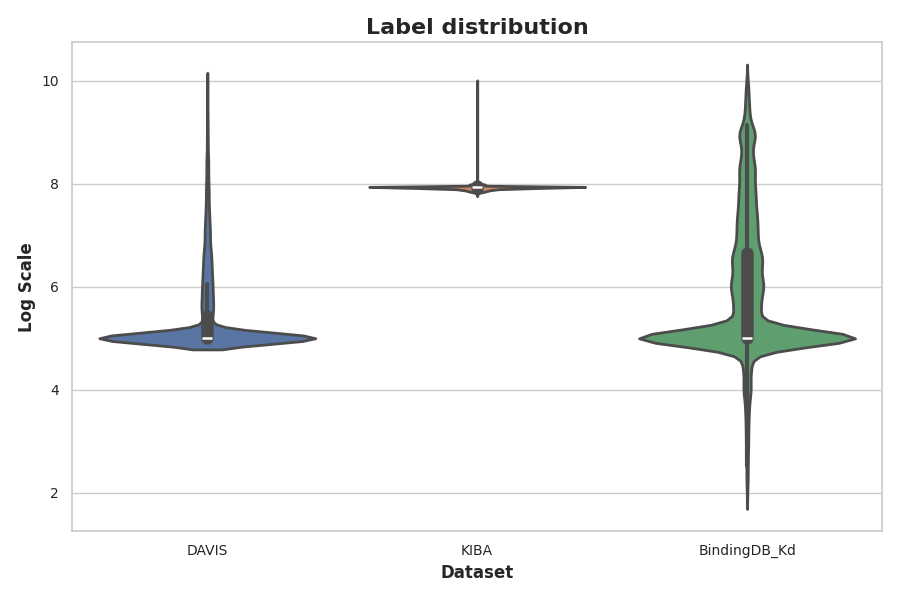}}
    \subfigure[Label distribution of Human, \textit{C. elegans} and Drugbank for classification tasks.]{
    \includegraphics[width=0.4\textwidth]{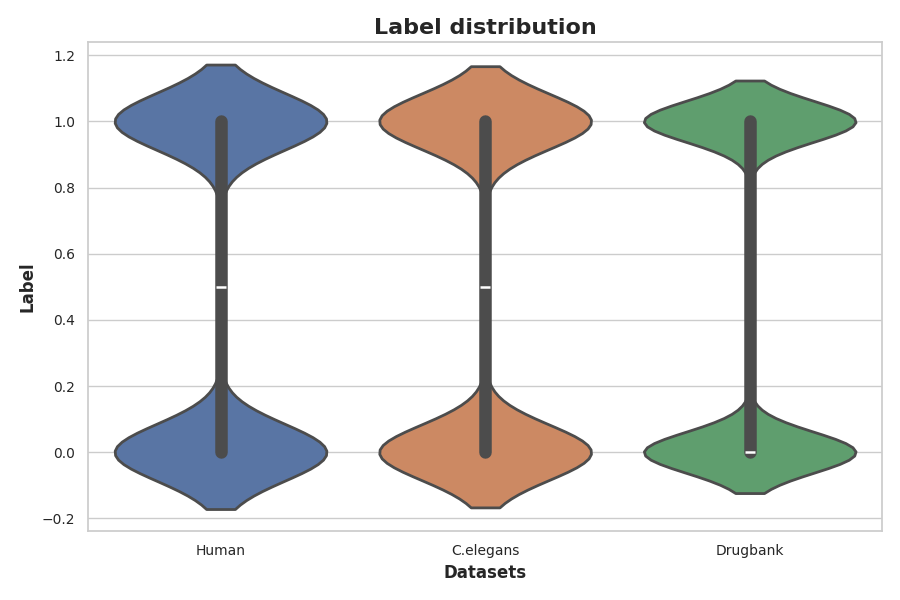}}
    \caption{Label distribution of different datasets for two tasks.}
    \label{fig:label_distribution}
\end{figure}
\newpage

\section{Evaluation Metrics}
\label{app:eva}
 We adopt distinct sets of metrics to evaluate the classification and regression tasks. In particular, considering the classification task, we utilize 
the common metrics, including Area Under Receiver Operating Characteristic Curve (ROC-AUC), Precision-Recall Area Under Curve (PR-AUC), LogAUC, accuracy, precision, recall, and F1 score.
For the continuous binding affinity regression, we 
benchmark the models using metrics of mean squared error (MSE), mean absolute error (MAE), coefficient of determination (R2), Pearson correlation coefficient, concordance index (CI), and Spearman correlation coefficient. Each of these metrics offers unique insights into different aspects of model performance, allowing us to assess predictive accuracy, correlation with observed values, and consistency in ranking predictions.

\newpage
\section{Original Hyperparameter} 
\label{sec:hyper}
To have a basic understanding of hyperparameters before greedy search and to find the optimized setting for each model, we summarize the hyperparameters reported in the corresponding paper or codes in Table \ref{tab:hyperparameters}.
\begin{table}[htbp]
\centering
\caption{Configurations of basic hyperparameters adopted to implement different approaches.}
\vspace{1em}
\resizebox{\textwidth}{!}{
\begin{tabular}{llcccccccccccc}
\toprule
Category & Models & Batch size & Total epoch & Learning rate \& Decay \& Decay epoch & Weight decay & Dropout & Optimizer \\
\midrule
\multirow{24}{*}{GNN} & GraphDTA-GCN \citep{nguyen2021graphdta}& 512 & 1000 & 0.0005 & - & 0.2 & Adam  \\
& GraphDTA-GAT \citep{nguyen2021graphdta}& 512 & 1000 & 0.0005 & - & 0.2 & Adam\\
& GraphDTA-GATGCN \citep{nguyen2021graphdta}& 512 & 1000 & 0.0005 & - & 0.2 & Adam\\
& GraphDTA-GIN \citep{nguyen2021graphdta} & 512 & 1000 & 0.0005 & - & 0.2 & Adam\\
& GraphCPI-GCN \citep{quan2019graphcpi} & 512 & 1000 & 0.0005 & - & 0.5 & Adam\\
& GraphCPI-GAT \citep{quan2019graphcpi}& 512 & 1000 & 0.0005 & - & 0.6 & Adam\\
& GraphCPI-GATGCN \citep{quan2019graphcpi}& 512 & 1000 & 0.0005 & - & - & Adam\\
& GraphCPI-GIN \citep{quan2019graphcpi}& 512 & 1000 & 0.0005 & - & 0.6 & Adam\\
& MGraphDTA \citep{yang2022MGraphDTA} & 512 & 3000 & 0.0005 & - & 0.1 & Adam \\
& SAGDTA \citep{zhang2021sag} & 512 & 2000 & 0.001 & - & 0.1 & Adam\\
& EmbedDTI \citep{jin2021embeddti} & 512 & 1500 & 0.0005 & - & 0.2 & Adam\\
& DeepGLSTM \citep{mukherjee2022deepglstm} & 512/128 & 1000 & 0.0005 & - & 0.2 & Adam\\
& CPI \citep{tsubaki2019compound} & 1 & 100 & 0.001, 0.5, 10 & 1e-6 & 0 & Adam\\
& BACPI \citep{li2022bacpi} & 16 & 20 & 0.0005, 0.5, 10 & - & 0.1 & Adam\\
& DeepNC-HGC \citep{tran2022deepnc} & 256 & 1000 & 0.0005 & - & 0.2 & Adam \\
& DeepNC-GEN \citep{tran2022deepnc}& 256 & 1000 & 0.0005 & - & 0.2 & Adam \\
& DrugBAN \citep{bai2023interpretable-drugban} & 64 & 100 & 0.00005 & - & 0 & Adam \\
& GANDTI \citep{wang2021gandti} & 1 & 30/15 & 0.001 & 1e-6 & 0.5 & Adam \\
& PGraphDTA-CNN \citep{bal2024pgraphdta} & 512 & 1500 & 0.0005 & - & 0.2 & Adam \\
& BridgeDPI \citep{wu2022bridgedpi} & 512 & 100 & 0.001 & - & 0.5 & Adam \\
& ColdDTA \citep{fang2023colddta} & 128 & 700/300 & 0.0003 & - & 0 & Adam \\
& SubMDTA \citep{pan2023submdta} & 512 & 1200 & 0.0005 & - & 0.2 & Adam \\
& IMAEN \citep{zhang2024imaen}& 128 & 1000 & 0.0005 & - & 0.2 & Adam \\ 
\midrule
\multirow{9}{*}{Transformer} 
& CSDTI \citep{pan2023csdti} & 256 & 3000 & 0.0005 & - &   0.2  & Adam\\
& AMMVF \citep{wang2023ammvf} & 32 & 40 & 0.001, 0.5, 5 & 1e-4 & 0.1 & Adam \\ 
& TDGraphDTA \citep{zhu2023tdgraph} & 1024 & 3000 & 0.0005 & - & 0.1 & Adam\\
& IIFDTI \citep{cheng2022iifdti} & 64 & 200 & 0.001 & 1e-6 & 0.2 & AdamW \\
& ICAN \citep{kurata2022ican} & 128 & 50 & 0.001 & - & 0.1 & Adam\\
& MolTrans \citep{huang2021moltrans} & 64 & 30 & 0.00001 & - & 0.1 & Adam\\
& TransformerCPI \citep{chen2020Transformercpi} & 8 & 40 & 0.0001, 0.5, 5 & 1e-4 & 0.2 & RAdam \\
& MRBDTA \citep{zhang2022predicting} & 1024/256 & 600/300 & 0.001 & - & 0.1 & Adam\\ 
& FOTFCPI \citep{yin2024fotf}& 64 & 100 & 0.0001 & - & 0.1 & Adam\\
\bottomrule
\end{tabular}
}
\label{tab:hyperparameters}
\end{table}

\newpage
\section{Full experiment}
\label{sec:expriment}
To evaluate the model's best performance, based on the hyperparameters given in its paper or codes, we found the optimized hyperparameters for each model. On top of the mean value, we also provide the standard deviation across five-fold. The complete result on the regression task is shown in Table \ref{tab:regression_bench_best}, and the complete result on the classification task is shown in Table \ref{tab:classification_bench_best}.

\begin{table}[ht]
\caption{Regression task benchmark on DAVIS, KIBA, and BindingDB datasets, respectively.}
\vspace{1em}
\centering
\resizebox{\textwidth}{!}{
\begin{tabular}{llccccccccccccccccccc}
\toprule
\multirow{2}{*}{Category} & \multirow{2}{*}{Models} & \multicolumn{6}{c}{\textbf{DAVIS}} & \multicolumn{6}{c}{\textbf{KIBA}} & \multicolumn{6}{c}{\textbf{BindingDB}}  \\
\cmidrule(r){3-8}  \cmidrule(lr){9-14} \cmidrule(lr){15-20} & 
&  MSE & MAE & R2 & PCC & CI & Spearman &  MSE ($\times 10^{-2}$) & MAE & R2 & PCC & CI & Spearman & MSE & MAE & R2 & PCC & CI & Spearman \\
\midrule
\multirow{24}{*}{GNN} 
& GraphDTA-GCN & $0.315 \pm 0.019$ & $0.332 \pm 0.017$ & $0.531 \pm 0.028$ & $0.734 \pm 0.018$ & $0.840 \pm 0.006$ & $0.605 \pm 0.010$ & $0.279 \pm 0.059$ & $0.041 \pm 0.005$ & $-2.114 \pm 0.662$ & $0.112 \pm 0.026$ & $0.555 \pm 0.014$ & $0.156 \pm 0.039$ & $0.609 \pm 0.033$ & $0.505 \pm 0.028$ & $0.669 \pm 0.018$ & $0.820 \pm 0.011$ & $0.837 \pm 0.007$ & $0.744 \pm 0.011$\\
& GraphDTA-GAT & $0.382 \pm 0.043$ & $0.380 \pm 0.025$ & $0.431 \pm 0.064$ & $0.671 \pm 0.039$ & $0.828 \pm 0.014$ & $0.588 \pm 0.024$ & $0.428 \pm 0.154$ & $0.051 \pm 0.011$ & $-3.769 \pm 1.712$ & $0.073 \pm 0.061$ & $0.534 \pm 0.026$ & $0.098 \pm 0.074$ & $1.020 \pm 0.032$ & $0.676 \pm 0.005$ & $0.445 \pm 0.018$ & $0.707 \pm 0.035$ & $0.786 \pm 0.017$ & $0.653 \pm 0.031$\\
& GraphDTA-GATGCN & $0.306 \pm 0.011$ & $0.325 \pm 0.013$ & $0.544 \pm 0.017$ & $0.741 \pm 0.013$ & $0.847 \pm 0.003$ & $0.617 \pm 0.006$ & $0.311 \pm 0.234$ & $0.041 \pm 0.018$ & $-2.467 \pm 2.604$ & $0.197 \pm 0.254$ & $0.578 \pm 0.094$ & $0.210 \pm 0.240$ & $0.574 \pm 0.032$ & $0.478 \pm 0.027$ & $0.687 \pm 0.017$ & $0.831 \pm 0.011$ & $0.844 \pm 0.007$ & $0.756 \pm 0.012$\\
& GraphDTA-GIN & $0.253 \pm 0.010$ & $0.295 \pm 0.014$ & $0.623 \pm 0.015$ & $0.791 \pm 0.008$ & $0.861 \pm 0.006$ & $0.638 \pm 0.009$ & $0.255 \pm 0.007$ & $0.039 \pm 0.006$ & $-1.840 \pm 0.779$ & $0.124 \pm 0.037$ & $0.553 \pm 0.019$ & $0.149 \pm 0.052$ & $0.563 \pm 0.038$ & $0.494 \pm 0.023$ & $0.693 \pm 0.021$ & $0.836 \pm 0.012$ & $0.842 \pm 0.007$ & $0.756 \pm 0.011$\\
& GraphCPI-GCN & $0.394 \pm 0.046$ & $0.401 \pm 0.025$ & $0.414 \pm 0.068$ & $0.652 \pm 0.051$ & $0.806 \pm 0.016$ & $0.550 \pm 0.027$ & $1.185 \pm 1.720$ & $0.066 \pm 0.057$ & $-12.202 \pm 19.159$ & $0.228 \pm 0.253$ & $0.602 \pm 0.104$ & $0.272 \pm 0.270$ & $1.199 \pm 0.040$ & $0.756 \pm 0.019$ & $0.347 \pm 0.022$ & $0.606 \pm 0.026$ & $0.727 \pm 0.015$ & $0.533 \pm 0.033$\\
& GraphCPI-GAT & $0.612 \pm 0.038$ & $0.501 \pm 0.021$ & $0.089 \pm 0.056$ & $0.419 \pm 0.135$ & $0.718 \pm 0.069$ & $0.396 \pm 0.123$ & $4.558 \pm 2.116$ & $0.168 \pm 0.048$ & $-49.762 \pm 23.563$ & $-0.021 \pm 0.023$ & $0.494 \pm 0.013$ & $0.031 \pm 0.024$ & $1.199 \pm 0.040$ & $0.756 \pm 0.019$ & $0.347 \pm 0.022$ & $0.606 \pm 0.026$ & $0.727 \pm 0.015$ & $0.533 \pm 0.033$\\
& GraphCPI-GATGCN & $0.338 \pm 0.013$ & $0.364 \pm 0.007$ & $0.496 \pm 0.019$ & $0.708 \pm 0.012$ & $0.838 \pm 0.004$ & $0.604 \pm 0.007$ & $0.445 \pm 0.081$ & $0.051 \pm 0.005$ & $-3.957 \pm 0.907$ & $0.063 \pm 0.050$ & $0.522 \pm 0.025$ & $0.081 \pm 0.045$ & $0.629 \pm 0.012$ & $0.520 \pm 0.011$ & $0.657 \pm 0.007$ & $0.813 \pm 0.005$ & $0.834 \pm 0.004$ & $0.739 \pm 0.007$\\
& GraphCPI-GIN & $0.274 \pm 0.009$ & $0.331 \pm 0.007$ & $0.593 \pm 0.013$ & $0.773 \pm 0.008$ & $0.851 \pm 0.008$ & $0.622 \pm 0.013$ & $1.681 \pm 0.946$ & $0.091 \pm 0.042$ & $-17.724 \pm 10.533$ & $0.142 \pm 0.220$ & $0.553 \pm 0.094$ & $0.149 \pm 0.246$ & $0.557 \pm 0.017$ & $0.475 \pm 0.016$ & $0.696 \pm 0.009$ & $0.838 \pm 0.006$ & $0.847 \pm 0.003$ & $0.760 \pm 0.006$\\
& MGraphDTA & $0.232 \pm 0.012$ & $0.268 \pm 0.008$ & $0.655 \pm 0.018$ & $0.812 \pm 0.011$ & $0.869 \pm 0.007$ & $0.650 \pm 0.011$ & $0.032 \pm 0.012$ & $0.011 \pm 0.002$ & $0.642 \pm 0.133$ & $0.803 \pm 0.079$ & $0.832 \pm 0.040$ & $0.793 \pm 0.070$ & $0.529 \pm 0.011$  & $0.444 \pm 0.025$ & $0.712 \pm 0.006$ & $0.847 \pm 0.005$ & $0.852 \pm 0.005$ & $0.769 \pm 0.008$\\
& SAGDTA & $0.324 \pm 0.064$ & $0.329 \pm 0.041$ & $0.518 \pm 0.096$ & $0.723 \pm 0.065$ & $0.833 \pm 0.027$ & $0.594 \pm 0.044$ & $0.065 \pm 0.008$ & $0.017 \pm 0.002$ & $0.279 \pm 0.085$ & $0.541 \pm 0.085$ & $0.713 \pm 0.032$ & $0.561 \pm 0.075$& $0.529 \pm 0.011$ & $0.444 \pm 0.025$ & $0.712 \pm 0.006$ & $0.847 \pm 0.005$ & $0.852 \pm 0.005$ & $0.769 \pm 0.008$ \\
& EmbedDTI & $0.280 \pm 0.024$ & $0.310 \pm 0.028$ & $0.583 \pm 0.036$ & $0.764 \pm 0.023$ & $0.851 \pm 0.009$ & $0.623 \pm 0.013$ & $0.289 \pm 0.142$ & $0.041 \pm 0.012$ & $-2.217 \pm 1.579$ & $0.131 \pm 0.090$ & $0.558 \pm 0.038$ & $0.164 \pm 0.106$ & $0.542 \pm 0.019$ & $0.446 \pm 0.021$ & $0.705 \pm 0.010$ & $0.843 \pm 0.006$ & $0.850 \pm 0.004$ & $0.767 \pm 0.007$\\
& DeepGLSTM & $0.316 \pm 0.023$ & $0.322 \pm 0.024$ & $0.529 \pm 0.035$ & $0.732 \pm 0.022$ & $0.841 \pm 0.007$ & $0.609 \pm 0.011$ & $8.539 \pm 7.479$ & $0.243 \pm 0.155$ & $-94.109 \pm 83.400$ & $0.040 \pm 0.083$ & $0.514 \pm 0.036$ & $0.071 \pm 0.078$ & $0.594 \pm 0.061$ & $0.474 \pm 0.046$ & $0.677 \pm 0.033$ & $0.825 \pm 0.021$ & $0.840 \pm 0.013$ & $0.750 \pm 0.021$\\
& CPI & $0.402 \pm 0.082$ & $0.393 \pm 0.054$ & $0.401 \pm 0.122$ & $0.629 \pm 0.101$ & $0.811 \pm 0.033$ & $0.560 \pm 0.055$ & $0.052 \pm 0.003$ & $0.161 \pm 0.008$ & $0.416 \pm 0.036$ & $0.654 \pm 0.032$ & $0.734 \pm 0.037$ & $0.605 \pm 0.088$ & $0.762 \pm 0.165$ & $0.565 \pm 0.068$ & $0.585 \pm 0.090$ & $0.768 \pm 0.063$ & $0.815 \pm 0.028$ & $0.704 \pm 0.054$\\
& BACPI & $0.334 \pm 0.015$ & $0.323 \pm 0.034$ & $0.502 \pm 0.023$ & $0.717 \pm 0.014$ & $0.827 \pm 0.006$ & $0.584 \pm 0.010$ & $0.031 \pm 0.004$ & $0.011 \pm 0.001$ & $0.658 \pm 0.043$ & $0.820 \pm 0.019$ & $0.831 \pm 0.020$ & $0.798 \pm 0.032$ & $0.550 \pm 0.010$ & $0.436 \pm 0.005$ & $0.700 \pm 0.006$ & $0.839 \pm 0.003$ & $0.845 \pm 0.002$ & $0.759 \pm 0.003$\\
& DeepNC-HGC & $0.309 \pm 0.025$ & $0.331 \pm 0.022$ & $0.541 \pm 0.037$ & $0.738 \pm 0.025$ & $0.841 \pm 0.005$ & $0.608 \pm 0.008$ & $0.080 \pm 0.003$ & $0.019 \pm 0.001$ & $0.110 \pm 0.036$ & $0.342 \pm 0.041$ & $0.667 \pm 0.022$ & $0.448 \pm 0.048$ & $0.572 \pm 0.011$ & $0.486 \pm 0.010$ & $0.689 \pm 0.006$ & $0.833 \pm 0.005$ & $0.844 \pm 0.003$ & $0.757 \pm 0.005$\\
& DeepNC-GEN & $0.270 \pm 0.012$ & $0.298 \pm 0.012$ & $0.597 \pm 0.017$ & $0.776 \pm 0.012$ & $0.852 \pm 0.009$ & $0.624 \pm 0.014$ & $0.135 \pm 0.045$ & $0.027 \pm 0.006$ & $-0.509 \pm 0.505$ & $0.266 \pm 0.073$ & $0.608 \pm 0.037$ & $0.301 \pm 0.097$ & $0.578 \pm 0.020$ & $0.474 \pm 0.012$ & $0.685 \pm 0.011$ & $0.830 \pm 0.006$ & $0.840 \pm 0.003$ & $0.749 \pm 0.005$\\
& DrugBAN & $0.242 \pm 0.007$ & $0.272 \pm 0.007$ & $0.640 \pm 0.010$ & $0.801 \pm 0.007$ & $0.869 \pm 0.003$ & $0.651 \pm 0.005$ & $0.029 \pm 0.003$ & $0.011 \pm 0.001$ & $0.676\pm0.032$ & $0.826\pm0.020$ & $0.832\pm0.013$ & $0.800\pm0.022$ & $0.465 \pm 0.018$ & $0.420 \pm 0.016$ & $0.747 \pm 0.010$ & $0.865 \pm 0.006$ & $0.862 \pm 0.003$ & $0.787 \pm 0.006$\\
& GANDTI & $0.318 \pm 0.018$ & $0.301 \pm 0.021$ & $0.527 \pm 0.027$ & $0.732 \pm 0.016$ & $0.844 \pm 0.006$ & $0.616 \pm 0.013$ & $0.030 \pm 0.002$ & $0.011 \pm 0.000$ & $0.662 \pm 0.026$ & $0.816 \pm 0.016$ & $0.831 \pm 0.007$ & $0.800 \pm 0.011$ & $0.621 \pm 0.012$ & $0.489 \pm 0.007$ & $0.662 \pm 0.006$ & $0.815 \pm 0.003$ & $0.836 \pm 0.002$ & $0.741 \pm 0.005$\\
& BridgeDPI & $1.241 \pm 1.432$ & $0.705 \pm 0.600$ & $-0.848 \pm 2.133$ & $0.657 \pm 0.209$ & $0.827 \pm 0.078$ & $0.581 \pm 0.128$ & $0.325 \pm 0.109$ & $0.010 \pm 0.000$ & $0.638 \pm 0.121$ & $0.821 \pm 0.038$ & $0.857 \pm 0.001$ & $0.839 \pm 0.002$ & $0.514 \pm 0.011$ & $0.413 \pm 0.006$ & $0.720 \pm 0.006$ & $0.853 \pm 0.003$ & $0.861 \pm 0.002$ & $0.783 \pm 0.003$\\
& ColdDTA & $0.220 \pm 0.009$ & $0.259 \pm 0.007$ & $0.672 \pm 0.014$ & $0.820 \pm 0.009$ & $0.880 \pm 0.004$ & $0.666 \pm 0.006$ & $0.110 \pm 0.029$ & $0.026 \pm 0.003$ & $-0.224 \pm 0.329$ & $0.441 \pm 0.217$ & $0.673 \pm 0.079$ & $0.459 \pm 0.191$ & $0.463 \pm 0.008$ & $0.391 \pm 0.007$ & $0.748 \pm 0.004$ & $0.866 \pm 0.002$ & $0.866 \pm 0.001$ & $0.789 \pm 0.002$\\
& SubMDTA & $0.289 \pm 0.012$ & $0.353 \pm 0.020$ & $0.570 \pm 0.018$ & $0.766 \pm 0.012$ & $0.841 \pm 0.007$ & $0.604 \pm 0.012$ & $0.029 \pm 0.002$ & $0.011 \pm 0.001$ & $0.678 \pm 0.025$ & $0.825 \pm 0.015$ & $0.836 \pm 0.011$ & $0.805 \pm 0.018$  & $0.532 \pm 0.032$ & $0.476 \pm 0.026$ & $0.710 \pm 0.017$ & $0.845 \pm 0.010$ & $0.852 \pm 0.006$ & $0.772 \pm 0.011$\\
& IMAEN & $0.230 \pm 0.009$ & $0.245 \pm 0.004$ & $0.657 \pm 0.014$ & $0.812 \pm 0.008$ & $0.874 \pm 0.004$ & $0.658 \pm 0.007$ & $0.046 \pm 0.018$ & $0.014 \pm 0.003$ & $0.484 \pm 0.196$ & $0.684 \pm 0.176$ & $0.781 \pm 0.056$ & $0.697 \pm 0.121$ & $0.479 \pm 0.012$ & $0.399 \pm 0.008$ & $0.739 \pm 0.006$ & $0.861 \pm 0.004$ & $0.863 \pm 0.002$ & $0.788 \pm 0.005$\\ 
\midrule
\multirow{10}{*}{Transformer} 
& CSDTI & $0.331 \pm 0.012$ & $0.339 \pm 0.020$ & $0.508 \pm 0.017$ & $0.720 \pm 0.009$ & $0.832 \pm 0.005$ & $0.593 \pm 0.008$ & $0.088 \pm 0.004$ & $0.020 \pm 0.001$ & $0.014 \pm 0.041$ & $0.273 \pm 0.084$ & $0.628 \pm 0.047$ & $0.350 \pm 0.124$ & $0.768 \pm 0.021$ & $0.572 \pm 0.012$ & $0.582 \pm 0.012$ & $0.765 \pm 0.008$ & $0.805 \pm 0.004$ & $0.689 \pm 0.006$\\
& TDGraphDTA & $0.222 \pm 0.005$ & $0.265 \pm 0.007$ & $0.669 \pm 0.008$ & $0.820 \pm 0.004$ & $0.653 \pm 0.011$ & $0.871 \pm 0.007$ & 
$0.091 \pm 0.019$ & $ 0.022 \pm 0.003 $ & $ -0.009 \pm 0.209 $ & $ 0.330 \pm 0.117 $ & $ 0.327 \pm 0.125 $ & $ 0.619 \pm 0.046 $
& $0.497 \pm 0.016$ & $0.418 \pm 0.010$ & $0.729 \pm 0.009$ & $0.855 \pm 0.005$ & $0.777 \pm 0.005$ & $0.857 \pm 0.003$\\
& AMMVF & $0.377 \pm 0.030$ & $0.365 \pm 0.043$ & $0.439 \pm 0.044$ & $0.669 \pm 0.036$ & $0.815 \pm 0.005$&  $0.586 \pm 0.024$ & $0.075 \pm 0.019$ & $0.019\pm0.003$ & $0.161 \pm 0.221$ & $0.679 \pm 0.003$ & $0.603 \pm 0.141$ & $0.659 \pm 0.001$ & $0.682 \pm 0.015$ & $0.517 \pm 0.010$ & $0.628 \pm 0.008$ & $0.796\pm0.004$ & $0.825 \pm 0.002$ & $0.721 \pm 0.004$\\
& IIFDTI & $0.313 \pm 0.018$ & $0.378 \pm 0.039$ & $0.534 \pm 0.027$ & $0.754 \pm 0.008$ & $0.836 \pm 0.008$ & $0.598 \pm 0.013$ & $0.054 \pm 0.012$ & $0.015 \pm 0.001$ & $0.398 \pm 0.143$ & $0.691 \pm 0.050$ & $0.767 \pm 0.009$ & $0.679 \pm 0.021$ & $0.634 \pm 0.024$ & $0.527 \pm 0.020$ & $0.655 \pm 0.013$ & $0.820 \pm 0.009$ & $0.832 \pm 0.006$ & $0.737 \pm 0.009$ \\
& ICAN & $0.371 \pm 0.013$ & $0.359 \pm 0.007$ & $0.448 \pm 0.020$ & $0.681 \pm 0.010$ & $0.818 \pm 0.006$ & $0.582 \pm 0.009$ & $0.089 \pm 0.000$ & $0.021 \pm 0.000$ & $-2.052 \pm 0.000$ & - & $0.500 \pm 0.000$ & - & $0.747 \pm 0.031$ & $0.580 \pm 0.018$ & $0.593 \pm 0.017$ & $0.785 \pm 0.006$ & $0.813 \pm 0.004$ & $0.707 \pm 0.005$\\
& MolTrans & $0.410 \pm 0.136$ & $0.382 \pm 0.045$ & $0.390 \pm 0.202$ & $0.670 \pm 0.107$ & $0.812 \pm 0.039$ & $0.591 \pm 0.042$ & $4.314 \pm 2.290$ & $0.169 \pm 0.055$ & $-47.055 \pm 25.515$ & $0.093 \pm 0.053$ & $0.540 \pm 0.021$ & $0.112 \pm 0.058$ & $0.695 \pm 0.183$ & $0.523 \pm 0.063$ & $0.621 \pm 0.100$ & $0.803 \pm 0.053$ & $0.822 \pm 0.009$ & $0.745 \pm 0.030$\\
& TransformerCPI & $0.393 \pm 0.022$ & $0.445 \pm 0.043$ & $0.415 \pm 0.032$ & $0.695 \pm 0.018$ & $0.802 \pm 0.008$ & $0.542 \pm 0.015$ & $0.070 \pm 0.003$ & $0.019 \pm 0.001$ & $0.217 \pm 0.033$ & $0.779 \pm 0.006$ & $0.800 \pm 0.002$ & $0.742 \pm 0.004$ & $0.659 \pm 0.040$ & $0.548 \pm 0.024$ & $0.641 \pm 0.022$ & $0.825 \pm 0.017$ & $0.829 \pm 0.013$ & $0.727 \pm 0.022$\\
& MRBDTA & $0.241 \pm 0.005$ & $0.265 \pm 0.006$ & $0.640 \pm 0.008$ & $0.802 \pm 0.003$ & $0.870 \pm 0.007$ & $0.651 \pm 0.011$ & $0.050 \pm 0.005$ & $0.016 \pm 0.001$ & $0.360 \pm 0.058$ & $0.600 \pm 0.050$ & $0.735 \pm 0.015$ & $0.613 \pm 0.031$ & $0.507 \pm 0.006$ & $0.411 \pm 0.006$ & $0.724 \pm 0.003$ & $0.853 \pm 0.002$ & $0.862 \pm 0.002$ & $0.786 \pm 0.003$\\ 
& FOTFCPI & $0.305 \pm 0.012$ & $0.302 \pm 0.019$ & $0.546 \pm 0.018$ & $0.749 \pm 0.011$ & $0.839 \pm 0.009$ & $0.604 \pm 0.015$ & $0.229 \pm 0.180$ & $0.034 \pm 0.016$ & $-1.554 \pm 2.003$ & $0.235 \pm 0.264$ & $0.587 \pm 0.086$ & $0.414 \pm 0.292$ & $0.567 \pm 0.008$ & $0.432 \pm 0.012$ & $0.695 \pm 0.004$ & $0.832 \pm 0.004$ & $0.848 \pm 0.006$ & $0.763 \pm 0.008$\\
& Our combos & $0.211 \pm 0.007$ & $0.251 \pm 0.008$ & $0.685 \pm 0.011$ & $0.829 \pm 0.006$ & $0.886 \pm 0.004$ & $0.676 \pm 0.007$ & $0.026 \pm 0.004$ & $0.010\pm 0.001$ & $0.710 \pm 0.051$ & $0.845 \pm 0.031$ & $0.849 \pm 0.023$ & $0.827 \pm 0.037$ & $0.461 \pm 0.006$ & $0.389 \pm 0.007$ & $0.749 \pm 0.003$ & $0.867 \pm 0.002$ & $0.869 \pm 0.002$ & $0.796 \pm 0.003$ \\
\bottomrule
\end{tabular}
}
\label{tab:regression_bench_best}
\end{table}

\begin{table}[ht]
\caption{Classification task benchmark on Human, \textit{C.elegans}, and Drugbank datasets, respectively.}
\vspace{1em}
\centering
\resizebox{\textwidth}{!}{
\begin{tabular}{llccccccccccccccccccccc}
\toprule
\multirow{2}{*}{Categories} & \multirow{2}{*}{Models} & \multicolumn{7}{c}{\textbf{Human}} & \multicolumn{7}{c}{\textbf{\textit{C.elegans}}} & \multicolumn{7}{c}{\textbf{Drugbank}}  \\
\cmidrule(r){3-9}  \cmidrule(lr){10-16} \cmidrule(lr){17-23} &
&  ROC-AUC & PR-AUC & Range-AUC & Acc. & Precision & Recall & F1 &  ROC-AUC & PR-AUC & Range-AUC & Acc. & Precision & Recall & F1 & ROC-AUC & PR-AUC & Range-AUC & Acc. & Precision & Recall & F1 \\
\midrule
\multirow{24}{*}{GNN}
& GraphDTA-GCN & $0.959\pm0.002$ & $0.950\pm0.002$ & $0.542\pm0.023$ & $0.898\pm0.005$ & $0.887\pm0.016$ & $0.884\pm0.011$ & $0.886\pm0.004$ & $0.974\pm0.002$ & $0.959\pm0.004$ & $0.587\pm0.028$ & $0.926\pm0.003$ & $0.911\pm0.017$ & $0.910\pm0.020$ & $0.910\pm0.004$ & $0.798\pm0.013$ & $0.757\pm0.024$ & $0.116\pm0.046$ & $0.751\pm0.007$ & $0.737\pm0.011$ & $0.778\pm0.005$ & $0.757\pm0.005$ \\
& GraphDTA-GAT & $0.947\pm0.004$ & $0.938\pm0.006$ & $0.499\pm0.042$ & $0.882\pm0.008$ & $0.855\pm0.018$ & $0.884\pm0.012$ & $0.869\pm0.007$ & $0.965\pm0.005$ & $0.952\pm0.007$ & $0.564\pm0.035$ & $0.909\pm0.011$ & $0.887\pm0.026$ & $0.894\pm0.015$ & $0.890\pm0.012$ & $0.804\pm0.005$ & $0.788\pm0.013$ & $0.145\pm0.028$ & $0.751\pm0.003$ & $0.734\pm0.009$ & $0.784\pm0.012$ & $0.758\pm0.001$ \\
& GraphDTA-GATGCN & $0.960\pm0.003$ & $0.952\pm0.007$ & $0.575\pm0.039$ & $0.907\pm0.010$ & $0.898\pm0.018$ & $0.892\pm0.017$ & $0.895\pm0.011$ & $0.978\pm0.003$ & $0.967\pm0.005$ & $0.641\pm0.053$ & $0.928\pm0.005$ & $0.909\pm0.011$ & $0.916\pm0.011$ & $0.913\pm0.006$ & $0.817\pm0.008$ & $0.787\pm0.014$ & $0.148\pm0.046$ & $0.765\pm0.004$ & $0.749\pm0.003$ & $0.795\pm0.007$ & $0.771\pm0.005$ \\
& GraphDTA-GIN & $0.949\pm0.007$ & $0.936\pm0.007$ & $0.491\pm0.038$ & $0.885\pm0.011$ & $0.879\pm0.026$ & $0.859\pm0.011$ & $0.869\pm0.011$ & $0.977\pm0.003$ & $0.970\pm0.004$ & $0.671\pm0.058$ & $0.929\pm0.005$ & $0.910\pm0.016$ & $0.919\pm0.012$ & $0.915\pm0.005$ & $0.850\pm0.001$ & $0.845\pm0.003$ & $0.246\pm0.038$ & $0.783\pm0.006$ & $0.776\pm0.011$ & $0.796\pm0.011$ & $0.785\pm0.005$ \\
& GraphCPI-GCN & $0.949\pm0.004$ & $0.934\pm0.008$ & $0.478\pm0.047$ & $0.890\pm0.003$ & $0.865\pm0.016$ & $0.893\pm0.024$ & $0.879\pm0.005$ & $0.968\pm0.004$ & $0.949\pm0.006$ & $0.535\pm0.041$ & $0.913\pm0.010$ & $0.898\pm0.024$ & $0.891\pm0.016$ & $0.894\pm0.011$ & $0.772\pm0.007$ & $0.725\pm0.011$ & $0.099\pm0.010$ & $0.739\pm0.005$ & $0.717\pm0.008$ & $0.788\pm0.003$ & $0.750\pm0.004$ \\
& GraphCPI-GAT & $0.921\pm0.003$ & $0.903\pm0.006$ & $0.364\pm0.022$ & $0.859\pm0.007$ & $0.817\pm0.015$ & $0.878\pm0.014$ & $0.847\pm0.006$ & $0.935\pm0.008$ & $0.909\pm0.010$ & $0.417\pm0.032$ & $0.876\pm0.006$ & $0.834\pm0.011$ & $0.871\pm0.015$ & $0.852\pm0.008$ & $0.756\pm0.006$ & $0.723\pm0.009$ & $0.082\pm0.008$ & $0.709\pm0.006$ & $0.678\pm0.007$ & $0.795\pm0.011$ & $0.732\pm0.006$ \\
& GraphCPI-GATGCN & $0.955\pm0.004$ & $0.944\pm0.005$ & $0.523\pm0.043$ & $0.899\pm0.006$ & $0.881\pm0.025$ & $0.894\pm0.018$ & $0.887\pm0.005$ & $0.970\pm0.002$ & $0.954\pm0.005$ & $0.553\pm0.030$ & $0.916\pm0.004$ & $0.900\pm0.029$ & $0.898\pm0.038$ & $0.898\pm0.007$ & $0.788\pm0.010$ & $0.745\pm0.015$ & $0.096\pm0.013$ & $0.748\pm0.005$ & $0.731\pm0.005$ & $0.783\pm0.013$ & $0.756\pm0.006$ \\
& GraphCPI-GIN & $0.941\pm0.005$ & $0.924\pm0.006$ & $0.427\pm0.028$ & $0.874\pm0.007$ & $0.860\pm0.016$ & $0.856\pm0.018$ & $0.858\pm0.008$ & $0.971\pm0.003$ & $0.957\pm0.006$ & $0.575\pm0.050$ & $0.924\pm0.008$ & $0.907\pm0.016$ & $0.908\pm0.014$ & $0.907\pm0.009$ & $0.838\pm0.012$ & $0.831\pm0.022$ & $0.214\pm0.051$ & $0.775\pm0.010$ & $0.765\pm0.017$ & $0.793\pm0.008$ & $0.778\pm0.006$ \\
& MGraphDTA & $0.960\pm0.004$ & $0.953\pm0.003$ & $0.542\pm0.030$ & $0.905\pm0.007$ & $0.889\pm0.019$ & $0.898\pm0.019$ & $0.893\pm0.007$ & $0.983\pm0.002$ & $0.976\pm0.004$ & $0.698\pm0.050$ & $0.943\pm0.004$ & $0.926\pm0.005$ & $0.935\pm0.004$ & $0.931\pm0.004$ & $0.879\pm0.004$ & $0.878\pm0.004$ & $0.303\pm0.023$ & $0.800\pm0.004$ & $0.782\pm0.016$ & $0.831\pm0.020$ & $0.806\pm0.003$ \\
& SAGDTA & $0.957\pm0.005$ & $0.947\pm0.008$ & $0.510\pm0.059$ & $0.901\pm0.005$ & $0.901\pm0.018$ & $0.874\pm0.021$ & $0.887\pm0.006$ & $0.966\pm0.006$ & $0.956\pm0.010$ & $0.603\pm0.064$ & $0.912\pm0.014$ & $0.883\pm0.021$ & $0.904\pm0.020$ & $0.894\pm0.017$ & $0.819\pm0.009$ & $0.809\pm0.010$ & $0.161\pm0.010$ & $0.752\pm0.010$ & $0.744\pm0.016$ & $0.770\pm0.019$ & $0.756\pm0.010$ \\
& EmbedDTI & $0.958\pm0.003$ & $0.947\pm0.003$ & $0.550\pm0.010$ & $0.901\pm0.005$ & $0.891\pm0.018$ & $0.886\pm0.018$ & $0.888\pm0.006$ & $0.975\pm0.003$ & $0.965\pm0.005$ & $0.641\pm0.056$ & $0.924\pm0.002$ & $0.899\pm0.007$ & $0.918\pm0.009$ & $0.908\pm0.002$ & $0.815\pm0.007$ & $0.785\pm0.014$ & $0.136\pm0.043$ & $0.758\pm0.005$ & $0.744\pm0.014$ & $0.787\pm0.018$ & $0.765\pm0.003$ \\
& DeepGLSTM & $0.958\pm0.004$ & $0.950\pm0.008$ & $0.571\pm0.044$ & $0.903\pm0.007$ & $0.891\pm0.018$ & $0.890\pm0.015$ & $0.890\pm0.008$ & $0.975\pm0.004$ & $0.963\pm0.007$ & $0.619\pm0.048$ & $0.923\pm0.006$ & $0.906\pm0.014$ & $0.907\pm0.013$ & $0.906\pm0.007$ & $0.796\pm0.014$ & $0.757\pm0.022$ & $0.145\pm0.048$ & $0.745\pm0.007$ & $0.730\pm0.010$ & $0.775\pm0.012$ & $0.752\pm0.006$ \\
& CPI & $0.951\pm0.012$ & $0.948\pm0.012$ & $0.579\pm0.078$ & $0.900\pm0.010$ & $0.890\pm0.013$ & $0.884\pm0.018$ & $0.887\pm0.012$ & $0.955\pm0.005$ & $0.944\pm0.011$ & $0.547\pm0.079$ & $0.913\pm0.007$ & $0.904\pm0.001$ & $0.881\pm0.019$ & $0.893\pm0.010$ & $0.739\pm0.087$ & $0.756\pm0.074$ & $0.162\pm0.069$ & $0.678\pm0.072$ & $0.666\pm0.064$ & $0.712\pm0.097$ & $0.687\pm0.074$ \\
& BACPI & $0.947\pm0.003$ & $0.938\pm0.005$ & $0.478\pm0.045$ & $0.905\pm0.007$ & $0.889\pm0.014$ & $0.898\pm0.010$ & $0.893\pm0.008$ & $0.975\pm0.003$ & $0.967\pm0.007$ & $0.638\pm0.077$ & $0.936\pm0.005$ & $0.932\pm0.009$ & $0.910\pm0.010$ & $0.921\pm0.006$ & $0.849\pm0.004$ & $0.836\pm0.005$ & $0.289\pm0.010$ & $0.776\pm0.009$ & $0.762\pm0.013$ & $0.803\pm0.016$ & $0.782\pm0.008$ \\
& DeepNC-HGC & $0.932\pm0.009$ & $0.913\pm0.010$ & $0.420\pm0.036$ & $0.861\pm0.015$ & $0.838\pm0.019$ & $0.853\pm0.014$ & $0.845\pm0.016$ & $0.970\pm0.003$ & $0.947\pm0.007$ & $0.477\pm0.025$ & $0.918\pm0.004$ & $0.881\pm0.023$ & $0.927\pm0.032$ & $0.903\pm0.006$ & $0.809\pm0.006$ & $0.777\pm0.006$ & $0.145\pm0.032$ & $0.752\pm0.006$ & $0.731\pm0.009$ & $0.795\pm0.011$ & $0.762\pm0.006$ \\
& DeepNC-GEN & $0.961\pm0.002$ & $0.954\pm0.003$ & $0.591\pm0.038$ & $0.907\pm0.006$ & $0.900\pm0.020$ & $0.890\pm0.020$ & $0.894\pm0.006$ & $0.980\pm0.002$ & $0.972\pm0.004$ & $0.703\pm0.060$ & $0.932\pm0.005$ & $0.916\pm0.009$ & $0.918\pm0.021$ & $0.917\pm0.008$ & $0.813\pm0.007$ & $0.803\pm0.018$ & $0.170\pm0.045$ & $0.736\pm0.015$ & $0.704\pm0.025$ & $0.817\pm0.025$ & $0.756\pm0.007$ \\
& DrugBAN & $0.974\pm0.002$ & $0.971\pm0.004$ & $0.688\pm0.050$ & $0.920\pm0.005$ & $0.905\pm0.011$ & $0.915\pm0.007$ & $0.910\pm0.005$ & $0.982\pm0.002$ & $0.974\pm0.008$ & $0.690\pm0.082$ & $0.946\pm0.004$ & $0.930\pm0.013$ & $0.940\pm0.012$ & $0.935\pm0.005$ & $0.876\pm0.004$ & $0.881\pm0.006$ & $0.303\pm0.028$ & $0.799\pm0.008$ & $0.788\pm0.021$ & $0.815\pm0.020$ & $0.801\pm0.005$ \\
& GANDTI & $0.970\pm0.002$ & $0.967\pm0.002$ & $0.676\pm0.033$ & $0.917\pm0.004$ & $0.913\pm0.010$ & $0.899\pm0.007$ & $0.906\pm0.004$ & $0.967\pm0.003$ & $0.963\pm0.004$ & $0.691\pm0.052$ & $0.919\pm0.007$ & $0.905\pm0.017$ & $0.897\pm0.008$ & $0.901\pm0.007$ & $0.836\pm0.014$ & $0.832\pm0.026$ & $0.226\pm0.038$ & $0.752\pm0.008$ & $0.730\pm0.017$ & $0.799\pm0.015$ & $0.763\pm0.004$ \\
& BridgeDPI & $0.957\pm0.012$ & $0.950\pm0.014$ & $0.564\pm0.065$ & $0.887\pm0.021$ & $0.849\pm0.042$ & $0.908\pm0.012$ & $0.877\pm0.020$ & $0.960\pm0.004$ & $0.943\pm0.010$ & $0.514\pm0.066$ & $0.882\pm0.034$ & $0.860\pm0.081$ & $0.869\pm0.101$ & $0.857\pm0.040$ & $0.726\pm0.076$ & $0.735\pm0.062$ & $0.138\pm0.028$ & $0.644\pm0.087$ & $0.632\pm0.095$ & $0.774\pm0.120$ & $0.685\pm0.047$ \\
& ColdDTA & $0.971\pm0.002$ & $0.967\pm0.003$ & $0.635\pm0.056$ & $0.922\pm0.009$ & $0.921\pm0.015$ & $0.903\pm0.013$ & $0.912\pm0.010$ & $0.983\pm0.003$ & $0.978\pm0.004$ & $0.728\pm0.059$ & $0.947\pm0.002$ & $0.937\pm0.011$ & $0.935\pm0.012$ & $0.936\pm0.002$ & $0.885\pm0.004$ & $0.884\pm0.006$ & $0.282\pm0.021$ & $0.813\pm0.005$ & $0.802\pm0.008$ & $0.830\pm0.002$ & $0.816\pm0.004$ \\
& SubMDTA & $0.971\pm0.003$ & $0.964\pm0.004$ & $0.610\pm0.045$ & $0.919\pm0.006$ & $0.911\pm0.017$ & $0.907\pm0.017$ & $0.909\pm0.007$ & $0.985\pm0.001$ & $0.981\pm0.002$ & $0.784\pm0.042$ & $0.945\pm0.007$ & $0.928\pm0.013$ & $0.939\pm0.007$ & $0.933\pm0.008$ & $0.861\pm0.005$ & $0.859\pm0.005$ & $0.269\pm0.013$ & $0.791\pm0.005$ & $0.783\pm0.010$ & $0.803\pm0.012$ & $0.793\pm0.005$ \\
& IMAEN & $0.944\pm0.004$ & $0.933\pm0.007$ & $0.478\pm0.052$ & $0.878\pm0.005$ & $0.862\pm0.021$ & $0.865\pm0.016$ & $0.863\pm0.003$ & $0.967\pm0.004$ & $0.956\pm0.006$ & $0.605\pm0.045$ & $0.911\pm0.007$ & $0.884\pm0.016$ & $0.901\pm0.011$ & $0.892\pm0.007$ & $0.847\pm0.004$ & $0.837\pm0.007$ & $0.218\pm0.028$ & $0.777\pm0.005$ & $0.766\pm0.012$ & $0.795\pm0.011$ & $0.780\pm0.004$ \\
\midrule
\multirow{10}{*}{Transformer}
& CSDTI & $0.905\pm0.007$ & $0.883\pm0.012$ & $0.339\pm0.035$ & $0.846\pm0.007$ & $0.831\pm0.009$ & $0.821\pm0.019$ & $0.826\pm0.009$ & $0.910\pm0.006$ & $0.877\pm0.010$ & $0.344\pm0.041$ & $0.840\pm0.011$ & $0.805\pm0.025$ & $0.805\pm0.010$ & $0.805\pm0.010$ & $0.774\pm0.011$ & $0.744\pm0.024$ & $0.096\pm0.029$ & $0.721\pm0.006$ & $0.705\pm0.009$ & $0.757\pm0.005$ & $0.730\pm0.004$ \\
& TDGraphDTA & $0.977\pm0.002$ & $0.973\pm0.003$ & $0.679\pm0.035$ & $0.927\pm0.005$ & $0.923\pm0.011$ & $0.911\pm0.013$ & $0.917\pm0.005$ & $0.984\pm0.001$ & $0.979\pm0.002$ & $0.739\pm0.041$ & $0.943\pm0.007$ & $0.950\pm0.008$ & $0.910\pm0.025$ & $0.929\pm0.010$ & $0.880\pm0.006$ & $0.882\pm0.008$ & $0.303\pm0.028$ & $0.805\pm0.006$ & $0.786\pm0.014$ & $0.836\pm0.011$ & $0.810\pm0.003$ \\
& AMMVF & $0.962\pm0.005$ & $0.957\pm0.008$ & $0.576\pm0.068$ & $0.915\pm0.007$ & $0.904\pm0.011$ & $0.906\pm0.024$ & $0.905\pm0.009$ & $0.984\pm0.005$ & $0.977\pm0.011$ & $0.721\pm0.130$ & $0.948\pm0.006$ & $0.941\pm0.018$ & $0.934\pm0.011$ & $0.937\pm0.007$ & $0.692\pm0.161$ & $0.695\pm0.149$ & $0.115\pm0.077$ & $0.654\pm0.088$ & $0.643\pm0.081$ & $0.784\pm0.121$ & $0.696\pm0.020$ \\
& IIFDTI & $0.973\pm0.006$ & $0.966\pm0.010$ & $0.647\pm0.057$ & $0.920\pm0.006$ & $0.916\pm0.017$ & $0.904\pm0.027$ & $0.909\pm0.008$ & $0.987\pm0.002$ & $0.981\pm0.004$ & $0.760\pm0.073$ & $0.948\pm0.005$ & $0.935\pm0.015$ & $0.939\pm0.010$ & $0.937\pm0.005$ & $0.849\pm0.014$ & $0.840\pm0.024$ & $0.204\pm0.045$ & $0.777\pm0.010$ & $0.763\pm0.007$ & $0.801\pm0.015$ & $0.782\pm0.011$ \\
& ICAN & $0.971\pm0.002$ & $0.966\pm0.006$ & $0.668\pm0.043$ & $0.927\pm0.005$ & $0.921\pm0.018$ & $0.913\pm0.013$ & $0.917\pm0.005$ & $0.977\pm0.004$ & $0.971\pm0.006$ & $0.709\pm0.044$ & $0.942\pm0.003$ & $0.940\pm0.010$ & $0.919\pm0.006$ & $0.929\pm0.004$ & $0.839\pm0.005$ & $0.841\pm0.004$ & $0.276\pm0.007$ & $0.764\pm0.005$ & $0.754\pm0.007$ & $0.781\pm0.005$ & $0.768\pm0.004$ \\
& MolTrans & $0.979\pm0.003$ & $0.975\pm0.005$ & $0.685\pm0.071$ & $0.931\pm0.002$ & $0.919\pm0.015$ & $0.927\pm0.016$ & $0.923\pm0.002$ & $0.980\pm0.003$ & $0.975\pm0.004$ & $0.715\pm0.053$ & $0.943\pm0.004$ & $0.939\pm0.018$ & $0.921\pm0.015$ & $0.930\pm0.004$ & $0.868\pm0.004$ & $0.873\pm0.003$ & $0.278\pm0.006$ & $0.795\pm0.005$ & $0.794\pm0.010$ & $0.796\pm0.026$ & $0.795\pm0.009$ \\
& TransformerCPI & $0.968\pm0.003$ & $0.958\pm0.006$ & $0.553\pm0.052$ & $0.917\pm0.004$ & $0.911\pm0.014$ & $0.901\pm0.010$ & $0.906\pm0.004$ & $0.984\pm0.001$ & $0.977\pm0.002$ & $0.699\pm0.018$ & $0.941\pm0.005$ & $0.918\pm0.014$ & $0.939\pm0.009$ & $0.929\pm0.006$ & $0.874\pm0.007$ & $0.877\pm0.008$ & $0.292\pm0.027$ & $0.799\pm0.008$ & $0.786\pm0.013$ & $0.820\pm0.011$ & $0.803\pm0.007$ \\
& MRBDTA & $0.971\pm0.004$ & $0.966\pm0.006$ & $0.625\pm0.076$ & $0.920\pm0.007$ & $0.916\pm0.021$ & $0.903\pm0.019$ & $0.909\pm0.007$ & $0.985\pm0.002$ & $0.983\pm0.002$ & $0.814\pm0.058$ & $0.953\pm0.002$ & $0.948\pm0.008$ & $0.939\pm0.007$ & $0.943\pm0.002$ & $0.866\pm0.005$ & $0.868\pm0.007$ & $0.273\pm0.023$ & $0.789\pm0.006$ & $0.781\pm0.013$ & $0.799\pm0.009$ & $0.790\pm0.004$ \\
& FOTFCPI & $0.980\pm0.003$ & $0.978\pm0.003$ & $0.718\pm0.037$ & $0.937\pm0.006$ & $0.928\pm0.017$ & $0.931\pm0.009$ & $0.929\pm0.006$ & $0.987\pm0.001$ & $0.984\pm0.003$ & $0.781\pm0.072$ & $0.953\pm0.003$ & $0.950\pm0.009$ & $0.935\pm0.009$ & $0.942\pm0.004$ & $0.866\pm0.002$ & $0.867\pm0.008$ & $0.261\pm0.029$ & $0.790\pm0.004$ & $0.780\pm0.007$ & $0.806\pm0.018$ & $0.793\pm0.006$ \\
& Our combos & $0.981\pm0.003$ & $0.979\pm0.003$ & $0.745\pm0.042$ & $0.936\pm0.007$ & $0.926\pm0.012$ & $0.929\pm0.007$ & $0.928\pm0.008$ & $0.987\pm0.003$ & $0.982\pm0.004$ & $0.732\pm0.056$ & $0.954\pm0.005$ & $0.950\pm0.019$ & $0.938\pm0.014$ & $0.944\pm0.006$ & $0.866\pm0.007$ & $0.862\pm0.016$ & $0.243\pm0.042$ & $0.793\pm0.009$ & $0.773\pm0.017$ & $0.825\pm0.014$ & $0.798\pm0.005$ \\
\bottomrule
\end{tabular}
}
\label{tab:classification_bench_best}
\end{table}

\newpage
\newpage

\clearpage
\section{Comparison of different featurization}
\label{sec:feat}
In this section, we present the summarized featurization methods in Table \ref{tab:featurize_table}, the detailed description of all properties is shown in Table \ref{tab:featurize_exp}. Besides, an ablation study on featurization strategies is in Table \ref{tab:featurize}.

\begin{table}[ht]
\centering
\caption{Summary of the featurization of the GNN-based model. Mol. Graphs means molecular graphs, and both mean using molecular graphs and fingerprints.}
\vspace{1em}
\resizebox{\textwidth}{!}{%
\begin{tabular}{@{}llccccccccccccccccc@{}}
\toprule
\multicolumn{2}{c}{\textbf{Model Information}} & \multicolumn{8}{c}{\textbf{Atomic Properties}} & \multicolumn{3}{c}{\textbf{Hydrogen Information}} & \multicolumn{2}{c}{\textbf{Electron Properties}} & \multicolumn{2}{c}{\textbf{Stereochemistry}} & \multicolumn{1}{c}{\textbf{Structure}} \\
\cmidrule(r){1-2} \cmidrule(lr){3-10} \cmidrule(lr){11-13} \cmidrule(lr){14-15} \cmidrule(l){16-17} \cmidrule(l){18-18}
Models & Graph & Atom Type & Degree  & Implicit Valence & Explicit Valence & Hybridization & Aromaticity & Formal Charge & \# Atom & \# Hs & \# Explicit Hs & \# Implicit Hs & \# Radical Electrons & Electron Affinity & CIP & Chirality & Ring \\
\midrule
GraphDTA & Mol. Graphs & \textcolor{blue}{\ding{51}} & \textcolor{blue}{\ding{51}} & \textcolor{blue}{\ding{51}} &  & & \textcolor{blue}{\ding{51}} & & & \textcolor{blue}{\ding{51}} & & & & & & & &  \\
GraphCPI & Mol. Graphs & \textcolor{blue}{\ding{51}} & \textcolor{blue}{\ding{51}} & \textcolor{blue}{\ding{51}} &  & & \textcolor{blue}{\ding{51}} & & & \textcolor{blue}{\ding{51}} & & & & & & & &  \\
MGraphDTA & Mol. Graphs & \textcolor{blue}{\ding{51}} & \textcolor{blue}{\ding{51}} & \textcolor{blue}{\ding{51}} & \textcolor{blue}{\ding{51}} & \textcolor{blue}{\ding{51}} & \textcolor{blue}{\ding{51}} & \textcolor{blue}{\ding{51}} & \textcolor{blue}{\ding{51}}& \textcolor{blue}{\ding{51}} & \textcolor{blue}{\ding{51}} & & \textcolor{blue}{\ding{51}} & \textcolor{blue}{\ding{51}} & & &  \\
SAGDTA & Mol. Graphs & \textcolor{blue}{\ding{51}} & \textcolor{blue}{\ding{51}} & \textcolor{blue}{\ding{51}} &  & & \textcolor{blue}{\ding{51}} & & & \textcolor{blue}{\ding{51}} & & & & & & & &  \\
EmbedDTI & Mol. Graphs & \textcolor{blue}{\ding{51}} & \textcolor{blue}{\ding{51}} & \textcolor{blue}{\ding{51}} & \textcolor{blue}{\ding{51}} & \textcolor{blue}{\ding{51}} & \textcolor{blue}{\ding{51}} &\textcolor{blue}{\ding{51}} & & \textcolor{blue}{\ding{51}} & & & & & & & \textcolor{blue}{\ding{51}} &  \\
DeepGLSTM & Mol. Graphs & \textcolor{blue}{\ding{51}} & \textcolor{blue}{\ding{51}} & \textcolor{blue}{\ding{51}} &  & & \textcolor{blue}{\ding{51}} & & & \textcolor{blue}{\ding{51}} & & & & & & & &  \\
CPI & Fingerprints & \textcolor{blue}{\ding{51}} & & & & & \textcolor{blue}{\ding{51}} & & & & & & & & & &   \\
BACPI & Fingerprints & \textcolor{blue}{\ding{51}} & & & & & \textcolor{blue}{\ding{51}} & & & & & & & & & & \\
DeepNC & Mol. Graphs & \textcolor{blue}{\ding{51}} & \textcolor{blue}{\ding{51}} & \textcolor{blue}{\ding{51}} &  & & \textcolor{blue}{\ding{51}} & & & \textcolor{blue}{\ding{51}} & & & & & & & & \\
DrugBAN & Mol. Graphs & \textcolor{blue}{\ding{51}} & \textcolor{blue}{\ding{51}} & \textcolor{blue}{\ding{51}} & & \textcolor{blue}{\ding{51}} & \textcolor{blue}{\ding{51}} & \textcolor{blue}{\ding{51}} & & \textcolor{blue}{\ding{51}} & & \textcolor{blue}{\ding{51}} & \textcolor{blue}{\ding{51}} & & & & \textcolor{blue}{\ding{51}}\\
GANDTI & Fingerprints & \textcolor{blue}{\ding{51}} & & & & & \textcolor{blue}{\ding{51}} & & & & & & & & & & & \\
PGraphDTA-CNN & Mol. Graphs & \textcolor{blue}{\ding{51}} & \textcolor{blue}{\ding{51}} & \textcolor{blue}{\ding{51}} & & \textcolor{blue}{\ding{51}} & \textcolor{blue}{\ding{51}} & \textcolor{blue}{\ding{51}}& & \textcolor{blue}{\ding{51}} &  & \textcolor{blue}{\ding{51}} & \textcolor{blue}{\ding{51}} & & & &\textcolor{blue}{\ding{51}}\\
BridgeDPI & Mol. Graphs& \textcolor{blue}{\ding{51}} & \textcolor{blue}{\ding{51}} & \textcolor{blue}{\ding{51}} & & \textcolor{blue}{\ding{51}} & \textcolor{blue}{\ding{51}} & \textcolor{blue}{\ding{51}} &\\
ColdDTA & Mol. Graphs & \textcolor{blue}{\ding{51}} & \textcolor{blue}{\ding{51}} & \textcolor{blue}{\ding{51}} & & \textcolor{blue}{\ding{51}}& \textcolor{blue}{\ding{51}} & &  & & && & & \textcolor{blue}{\ding{51}} 
&\textcolor{blue}{\ding{51}} & \\
SubMDTA & Mol. Graphs & \textcolor{blue}{\ding{51}} & \textcolor{blue}{\ding{51}} & \textcolor{blue}{\ding{51}} &  & & \textcolor{blue}{\ding{51}} & & & \textcolor{blue}{\ding{51}} & & & & & & & &  \\
IMAEN & Mol. Graphs & \textcolor{blue}{\ding{51}} & \textcolor{blue}{\ding{51}} & \textcolor{blue}{\ding{51}} &  & & \textcolor{blue}{\ding{51}} & & & \textcolor{blue}{\ding{51}} & & & & & & & &  \\
CSDTI & Mol. Graphs & \textcolor{blue}{\ding{51}} & \textcolor{blue}{\ding{51}} & \textcolor{blue}{\ding{51}} & \textcolor{blue}{\ding{51}} & \textcolor{blue}{\ding{51}} & \textcolor{blue}{\ding{51}} & \textcolor{blue}{\ding{51}} & \textcolor{blue}{\ding{51}}& \textcolor{blue}{\ding{51}} & \textcolor{blue}{\ding{51}} & & \textcolor{blue}{\ding{51}} & \textcolor{blue}{\ding{51}} & & &   \\
TDGraphDTA & Mol. Graphs & \textcolor{blue}{\ding{51}} & \textcolor{blue}{\ding{51}} & \textcolor{blue}{\ding{51}} &  & & \textcolor{blue}{\ding{51}} & & & \textcolor{blue}{\ding{51}} & & & & & & & &  \\
AMMVF & Both & \textcolor{blue}{\ding{51}} & \textcolor{blue}{\ding{51}} & & & \textcolor{blue}{\ding{51}} & \textcolor{blue}{\ding{51}} & & & \textcolor{blue}{\ding{51}} & & & \textcolor{blue}{\ding{51}} & & \textcolor{blue}{\ding{51}}  \\
TransformerCPI & Mol. Graphs &\textcolor{blue}{\ding{51}} & \textcolor{blue}{\ding{51}} & & & \textcolor{blue}{\ding{51}} & \textcolor{blue}{\ding{51}} & \textcolor{blue}{\ding{51}} & &  \textcolor{blue}{\ding{51}} & & & \textcolor{blue}{\ding{51}} & & \textcolor{blue}{\ding{51}} & \textcolor{blue}{\ding{51}}\\
\bottomrule
\end{tabular}
}
\label{tab:featurize_table}
\end{table}

\begin{table}[ht]
\centering
\caption{Description of atomic and molecular properties for node featurization}
\vspace{1em}
\resizebox{15cm}{!}{ 
\begin{tabular}{|>{\centering\arraybackslash}p{4cm}|>{\centering\arraybackslash}p{10cm}|}
\hline
\textbf{Name} & \textbf{Description} \\ \hline
\multicolumn{2}{|c|}{\textbf{Atomic Properties}} \\ \hline
Atom Type & Type of the atom (e.g., C, N, O, H) \\ \hline
Degree & Number of directly bonded neighbors \\ \hline
Implicit Valence & Number of implicit valence of the atom \\ \hline
Explicit Valence & Number of explicit valence of the atom \\ \hline
Hybridization & The state of hybridization (e.g., sp3, sp2) \\ \hline
Aromaticity & Whether the atom is part of an aromatic system \\ \hline
Formal Charge & The charge assigned to an atom \\ \hline
\# Atom & Total number of atoms \\ \hline
\multicolumn{2}{|c|}{\textbf{Hydrogen Information}} \\ \hline
\# Hs & Total number of hydrogens \\ \hline
\# Explicit Hs & Number of explicit hydrogens on the atom \\ \hline
\# Implicit Hs & Number of implicit hydrogens on the atom \\ \hline
\multicolumn{2}{|c|}{\textbf{Electron Properties}} \\ \hline
\# Radical Electrons & Number of radical electrons \\ \hline
Electron Affinity & Tendency of an atom to accept electrons \\ \hline
\multicolumn{2}{|c|}{\textbf{Stereochemistry}} \\ \hline
CIP & The CIP code (R or S) of the atom \\ \hline
Chirality & If an atom is a possible chiral center \\ \hline
\multicolumn{2}{|c|}{\textbf{Structure}} \\ \hline
Ring & Whether the atom is part of a ring structure \\ \hline
\end{tabular}
}
\label{tab:featurize_exp}
\end{table}

\begin{table}[ht]
\centering
\caption{Extra Graph embedding feature exploration. Here Basic: \{Atom Type, Degree, Implicit Valence, Aromaticity, \# Hs\}}
\vspace{1em}
\resizebox{\textwidth}{!}{
\begin{tabular}{llccccccccccccccc}
\toprule
\multirow{2}{*}{Models} & \multirow{2}{*}{Intial Feature} & \multicolumn{6}{c}{Regression} & \multicolumn{7}{c}{Classification} \\
  \cmidrule(lr){3-8} \cmidrule(lr){9-15}
&  & MSE & MAE & R2 & PCC & CI & Spearman & ROC-AUC & PR-AUC & Range-AUC & Acc. & Precision & Recall & F1 \\
\midrule
\multirow{10}{*}{GraphDTA} & Basic & 0.2771 & 0.2947 & 0.5873 & 0.7695 & 0.8521 & 0.6241 & 0.9170 & 0.8873 & 0.4830 & 0.9171 & 0.9204 & 0.9189 & 0.9194 \\
& Basic+AP & 0.2772 & 0.2978 & 0.5873 & 0.7671 & 0.8496 & 0.6200 & 0.9153 & 0.8817 & 0.4517 & 0.9157 & 0.9099 & 0.9290 & 0.9191\\
& Basic+HI & 0.2783 & 0.2983 & 0.5855 & 0.7663 & 0.8483 & 0.6185 & 0.9211 & 0.8903 & 0.4815 & 0.9214 & 0.9188 & 0.9296 & 0.9241\\
& Basic+EP & 0.2775 & 0.3068 & 0.5868 & 0.7682 & 0.8499 & 0.6205 & 0.9165 & 0.8862 & 0.4795 & 0.9166 & 0.9185 & 0.9198 & 0.9191\\
& Basic+Ste & 0.2838 & 0.3030 & 0.5773 & 0.7624 & 0.8523 & 0.6254 & 0.9200 & 0.8905 & 0.4869 & 0.9200 & 0.9216 & 0.9235 & 0.9224\\
& Basic+Str & 0.2783 & 0.2991 & 0.5857 & 0.7668 & 0.8505 & 0.6228 & 0.9198 & 0.8865 & 0.4649 & 0.9202 & 0.9124 & 0.9351 & 0.9235\\
& Basic+AP+HI & 0.2851 & 0.3029 & 0.5755 & 0.7610 & 0.8504 & 0.6222 & 0.9163 & 0.8822 & 0.4629 & 0.9168 & 0.9094 & 0.9313 & 0.9201\\
& Basic+AP+HI+EP & 0.2845 & 0.2917 & 0.5763 & 0.7620 & 0.8510 & 0.6227 & 0.9140 & 0.8811 & 0.4580 & 0.9143 & 0.9115 & 0.9232 & 0.9173\\
& Basic+AP+HI+EP+Ste & 0.2811 & 0.3099 & 0.5814 & 0.7640 & 0.8500 & 0.6212 & 0.9192 & 0.8899 & 0.4853 & 0.9193 & 0.9218 & 0.9215 & 0.9216\\
& Basic+AP+HI+EP+Ste+Str & 0.2801 & 0.2916 & 0.5829 & 0.7659 & 0.8538 & 0.6278 & 0.9217 & 0.8905 & 0.4794 & 0.9220 & 0.9180 & 0.9319 & 0.9248\\
\midrule
\multirow{10}{*}{GraphCPI} & Basic
& 0.3291 & 0.3388 & 0.5100 & 0.7265 & 0.8294 & 0.5885 & 0.9060 & 0.8706 & 0.4385 & 0.9064 & 0.9029 & 0.9169 & 0.9098 \\
& Basic+AP & 0.3331 & 0.3389 & 0.5040 & 0.7198 & 0.8223 & 0.5761 & 0.9038 & 0.8657 & 0.4103 & 0.9043 & 0.8955 & 0.9218 & 0.9084\\
& Basic+HI & 0.3402 & 0.3457 & 0.4934 & 0.7157 & 0.8228 & 0.5769 & 0.9051 & 0.8713 & 0.4495 & 0.9052 & 0.9058 & 0.9094 & 0.9080\\
& Basic+EP & 0.3408 & 0.3505 & 0.4926 & 0.7123 & 0.8211 & 0.5749 & 0.9053 & 0.8713 & 0.4442 & 0.9055 & 0.9060 & 0.9111 & 0.9085\\
& Basic+Ste & 0.3398 & 0.3634 & 0.4940 & 0.7119 & 0.8274 & 0.5855 & 0.9061 & 0.8692 & 0.4261 & 0.9065 & 0.8992 & 0.9221 & 0.9104\\
& Basic+Str & 0.3419 & 0.3562 & 0.4909 & 0.7113 & 0.8226 & 0.5766 & 0.9066 & 0.8683 & 0.4079 & 0.9073 & 0.8957 & 0.9281 & 0.9115\\
& Basic+AP+HI & 0.3326 & 0.3471 & 0.5048 & 0.7212 & 0.8210 & 0.5734 & 0.9010 & 0.8659 & 0.4288 & 0.9012 & 0.9018 & 0.9071 & 0.9043\\
& Basic+AP+HI+EP & 0.3404 & 0.3476 & 0.4931 & 0.7150 & 0.8212 & 0.5748 & 0.9015 & 0.8612 & 0.3821 & 0.9022 & 0.8890 & 0.9258 & 0.9070\\
& Basic+AP+HI+EP+Ste & 0.3403 & 0.3445 &0.4932 & 0.7111 & 0.8169 & 0.5671 & 0.9109 & 0.8763 & 0.4511 & 0.9113 & 0.9065 & 0.9229 & 0.9146\\
& Basic+AP+HI+EP+Ste+Str & 0.3469& 0.3550& 0.4834& 0.7073& 0.8228& 0.5775& 0.9134 & 0.8772 & 0.4440 & 0.9140 & 0.9033 & 0.9328 & 0.9178\\
\bottomrule
\end{tabular}
}
\label{tab:featurize}
\end{table}

\clearpage

\newpage
\section{Memory and Parameter Comparison} 

\label{app:memory}
\begin{table}[ht]
\caption{Training time per epoch (s) and the max allocated memory (MB) for representative datasets on both regression (Davis) and classification (Human) tasks when BS is 32.}
\vspace{1em}
\centering
\resizebox{\textwidth}{!}{
\begin{tabular}{lcccccccccccc}
\toprule
\multirow{2}{*}{Categories}& \multirow{2}{*}{Models}  & \multicolumn{3}{c}{\textbf{Regression}} & \multicolumn{3}{c}{\textbf{Classification}}  \\
\cmidrule(lr){3-5} \cmidrule(lr){6-8} 
 & &Model parameter & Memory Usage (MB) & Time(s)  & Model parameter & Memory Usage (MB) & Run Time (s) \\
\midrule
\multirow{24}{*}{Graph} & GraphDTA-GCN & 7.87 & 86.45 & 8.92 & 7.87 & 86.33 & 2.43  \\
& GraphDTA-GAT & 6.58 & 104.71 & 9.62 & 6.58 & 99.40 & 2.43\\
& GraphDTA-GATGCN & 18.12 & 148.25 & 8.37 & 18.12 & 145.13 & 2.35\\
& GraphDTA-GIN  & 5.97 &  78.00 & 12.33 & 5.95 & 77.47 & 3.13\\
& GraphCPI-GCN & 10.46 & 98.13 & 7.02 & 10.48 & 63.37& 1.92\\
& GraphCPI-GAT & 9.16 & 116.19 & 9.38 & 9.18 & 112.34 & 2.48\\
& GraphCPI-GATGCN & 20.70 & 158.21 & 9.47 & 20.73 & 156.22 & 2.20\\
& GraphCPI-GIN & 8.55 & 88.55 & 12.54 & 8.56 & 88.02 & 2.92\\
& MGraphDTA & 11.75 & 235.97 & 69.84 & 11.43 & 217.15 & 17.59 \\
& SAGDTA & 7.45 & 88.31 & 20.87 & 7.44 & 87.54 & 4.34\\
& EmbedDTI & 16.97 & 152.55 & 17.80 & 16.97 & - & -\\
& DeepGLSTM & 131.92 & 1287.92 & 20.69 & 131.93 & 1287.16 & 11.22\\
& CPI & 0.37 & 14.00 & 11.29 & 0.6 & 14.82 & 2.69\\
& BACPI & 4.05 & 1051.91 & 43.27 & 6.13 & 1058.95 & 12.38\\
& DeepNC-HGC & 16.61 & 123.70 & 9.85 & 16.60 & 123.65 & 3.46 \\
& DeepNC-GEN & 18.84 & 174.00 & 11.35 & 18.84 & 166.55 & 3.46 \\
& DrugBAN & 4.10 & 940.22 & 30.06 & 4.10 & 940.23 & 7.84 \\
& GANDTI & 1.48 & 35.89 & 6.01 & 2.43 & 39.95 & 1.54 \\
& PGraphDTA-CNN & 9.03 & 102.85 & 13.71 & 9.03 & - & - \\
& BridgeDPI & 39.32 & 232.53 & 16.27 & 39.32 & 232.53 & 4.36 \\
& ColdDTA & 13.14 & 282.74 & 72.98 & 13.14 & 262.91 & 18.56 \\
& SubMDTA & 169.37 & 992.61 & 35.12 & 195.50 & 1095.73 & 8.49 \\
& IMAEN & 10.43 & 174.34 & 35.77 & 10.43 & 172.86 & 4.41 \\ 
\midrule
\multirow{8}{*}{Transformer} 
& CSDTI & 9.67 & 281.23 & 17.66 & 9.66 &   281.02  & 4.35\\
& TDGraphDTA & 8.62 & 247.23& 116.38& 8.62 & 236.02& 28.43\\
& AMMVF & 6.68 & 17847.62 & 216.20 & 7.49 & 17850.79 & 57.99 \\ 
& IIFDTI & 10.75 & 7946.92 & 141.12 & 10.75 & 11890.79 & 56.95 \\
& ICAN & 63.89 & 649.55 & 12.44 & 63.89 & 648.67 & 2.84\\
& MolTrans & 239.73 & 10624.55 & 70.19 & 239.74 & 10624.55 & 25.06\\
& TransformerCPI & 4.44 & 1219.58 & 28.98 & 4.45 & 1219.60 & 7.17 \\
& MRBDTA & 17.83 & 3893.76 & 66.47 & 17.84 & 3893.78 & 16.13\\ 

& FOTFCPI & 189.15& 6780.35& 58.75& 189.15 & 6780.35 & 14.80\\
& Our &19.02&1081.99&94.71 &19.02 & 1082.68 & 13.68\\
\bottomrule
\end{tabular}
}
\label{tab:my_label}
\end{table}



\end{document}